\documentclass[12pt]{iopart}

\usepackage{graphicx}
\begin{document}

\title[Proving the existence of bound states \ldots]{Proving the existence of bound states for attractive potentials in 1-d and 2-d without calculus}

\author{J. Alexander Jacoby}

\address{Department of Physics, Box 1843, 182 Hope St., Brown University, Providence, Rhode Island 02912, USA}

\author{Maurice Curran}

\address{George C. Marshall High School, 7731 Leesburg Pike, Falls Church, Virginia, 22043, USA}

\author{David R. Wolf}

\address{Department of Physics, Astronomy, and Engineering, Austin Community College, 5930 Middle Fiskville Rd., Austin, Texas 78752 USA}

\author{James K. Freericks}

\address{Department of Physics, Georgetown University, 37th and O Sts., NW, Washington, DC 20057, USA}

\ead{james.freericks@georgetown.edu}
\vspace{10pt}
\begin{indented}
\item[]\today
\end{indented}

\begin{abstract}
Schr\"odinger developed an operator method for solving quantum mechanics. While this technique is overshadowed by his more familiar differential equation approach, it has found wide application as an illustration of supersymmetric quantum mechanics. One reason for the reticence in its usage for conventional quantum instruction is that the approach for simple problems like the particle-in-a-box is much more complicated than the differential equation approach, making it appear to be less useful for pedagogy. We argue that the approach is still quite attractive because it employs only algebraic methods, and thereby has a much lower level of math background needed to use it. We show how Schr\"odinger's operator method can be streamlined for these particle-in-a-box problems greatly reducing the complexity of the solution and making it much more accessible. As an application, we illustrate how this approach can be used to prove an important result, the existence of bound states for one- and two-dimensional attractive potentials, using only algebraic methods. The approach developed here can be employed in undergraduate classes and possibly even high school classes because it employs only algebra and requires essentially no calculus.
\end{abstract}

%
%
\submitto{\EJP}
%
\maketitle
%
%

\section{Introduction}

Most potentials that describe quantum systems are limited in their extent and are attractive over some region of space. 
In spite of our ability to solve these problems either analytically or numerically, we still do not fully understand simple general principles, such as how many bound states does a given potential have? For example, in one- and two-dimensions, it is well known that any (piecewise continuous) attractive potential [$V(x)\le 0$] supports at least one bound state. But potentials like the simple harmonic oscillator, or the Coulomb problem support an infinite number of bound states. Is there a methodology that allows us to determine how many bound states a potential can support? This is a hard problem, and some progress has been made~\cite{ttwu} by employing variational methods in concert with the node theorem. In this work, we focus on the simpler problem of proving the well-known result that attractive potentials in one- and two-dimensions always have at least one bound state. We employ the Schr\"odinger operator method and thereby achieve this goal {\it without using any calculus}. We believe this makes this important problem more accessible to teach and illustrates that in spite of common beliefs that calculus is necessary for quantum mechanics, it is not.

Fifteen years after Schr\"odinger wrote his famous paper that introduced the Schr\"odinger equation, he introduced an alternative method for performing quantum-mechanical calculations called the factorization method and based on the  algebraic manipulation of operators~\cite{schrodinger_op1,schrodinger_op2}, which was reviewed and extended in the 1950s~\cite{infield_hull}. While this methodology has not been employed much in quantum mechanics texts, it first appeared in Harris and Loeb's book~\cite{harris_loeb}. This was followed by an extended treatment in Green~\cite{green}, O'Hanian~\cite{ohanian}, Bohm~\cite{bohm} and Binney and Skinner~\cite{binney_skinner}. It has also appeared in more recent books~\cite{delange,hecht,schwabl,razavy}. The approach has also been adopted for supersymmetric quantum mechanics~\cite{supersym1,supersym2,supersym3}. In all of these treatments, it is assumed that the student is equally agile in both algebra and calculus. Our emphasis here, is to show that these methods can be implemented without employing calculus. Indeed, much of the quantum-mechanical curriculum can be developed in this fashion---one of us is working on a book to do just that~\cite{quantum_wo_calculus}. 

Schr\"odinger, and all of the above texts, solve only the particle in an infinite square-well potential. Recent work has described how one can apply this method to solve the problem in a finite square-well potential~\cite{brazil} or with delta-function potentials~\cite{iran}. Note, however, that these latter papers do not develop the theory along the lines of Schr\"odinger's infinite chain of auxiliary Hamiltonians, but rather solves these problems with one (or two) factorizations. This is because the original Schr\"odinger method cannot be extended to either particles in finite square-well boxes or to particles in delta function potentials.
We illustrate below how a simple generalization of the Schr\"odinger methodology allows us to construct all of the bound-state solutions from just one (or two) factorization(s) of the Hamiltonian! This greatly simplifies the implementation of the method and makes it accessible to even high school students. This methodology is similar to the differential equation-based approaches and can be easily adopted with students as a means of introducing quantum mechanics to them. While this new approach is hinted at in both references above~\cite{brazil,iran}, it is not fully developed there.

The existence of at least one bound state for attractive one-dimensional and two-dimensional systems has been known for a long time. We examined 70 quantum textbooks and found that only six texts discussed this problem with enough detail that one can actually derive the result, instead of merely stating it. Landau and Lifshitz~\cite{landau_lifshitz} provide a masterful argument about the existence of bound states in one and two dimensions, but it is a rather technical discussion. Park~\cite{park} appears to be the first of the modern texts to include a problem that employs Gaussian wavefunctions within a variational argument to show the existence of at least one bound state in one dimension. This problem also appears in Gasiorowicz's book~\cite{gasiorowicz}, Shankar's book~\cite{shankar}, and Commins book~\cite{commins}. Zelevinsky~\cite{zelevinsky} provides a rather detailed account about particle in finite square wells, but does not detail the proofs in one and two dimensions. Finally, Robinett's book~\cite{robinett} is the only text we found that presents the simple variational argument we employ here for the one-dimensional problem.  This argument is similar in spirit to the Gaussian wavefunction argument, but uses the fact that one can explicitly show a bound state for a particle in a box and then employ a variational argument for all other attractive potentials that can have any attractive finite square well potential drawn within them. We illustrate below how this simple argument can be employed for both one and two-dimensional systems.

In the remainder of the article, we apply this simplified methodology to one-dimensional boxes (particle in an infinite box and particle on a circle) in Sec. II, including a discussion of how this approach confronts some of the subtle and oft neglected issues one has with these solutions. Section III derives how to solve the particle in a finite one-dimensional box. Section IV develops the $m=0$ solutions for the particle on a circle and then generalizes those solutions to all nonzero integer $m$. In Sec.V, we discuss the particle in a three-dimensional box. Finally, we present the variational argument for the existence of bound states in Sec. VI and we conclude in Sec. VII. Five appendices provide technical details that would have interrupted the flow of the main arguments.

\section{Particle particle in a one-dimensional box and on a circle}

We start by showing how to employ the Schr\"odinger operator approach in a simpler fashion to solve the particle in a box. Here, the potential vanishes for $-L/2\le x\le L/2$ and is infinite elsewhere. We find the Hamiltonian $\mathcal H$ inside the box is simply
\begin{equation}
\hat\mathcal{H}=\frac{\hat p^2}{2M},
\label{eq: particle-in-box_ham}
\end{equation}
where $\hat p$ is the momentum operator and $M$ is the mass of the particle (we use ``hats'' to denote operators throughout). The momentum and position operators satisfy the canonical commutation relation $[\hat x,\hat p]=i\hbar$. The standard operator approach from Schr\"odinger is to note that we factorize the Hamiltonian into a product of raising and lowering operators plus a constant:
\begin{equation}
\hat\mathcal{H}=\hat A^\dagger\hat A+E_0,
\label{eq: factorized}
\end{equation}
where we choose the factorization with the largest $E_0$ if there is any ambiguity. Then, Schr\"odinger has an automated procedure to create a series of auxiliary Hamiltonians from which one finds the higher-energy eigenstates. We proceed somewhat differently. We also  factorize the Hamiltonian in the same fashion as Schr\"odinger does for the ground state. For the particle in a box, one finds
\begin{equation}
\hat A_k=\frac{1}{\sqrt{2M}}\left [\hat p-i\hbar k\tan (k\hat x)\right ],
\label{eq: pib_a}
\end{equation}
where we employ a label $k$ for the lowering operator. Now the operator is well defined as long as $k$ runs from 0 up to $\pi/L$, at which point the lowering operator will diverge at the points where $x=\pm L/2$. Evaluating the operator $\hat A^\dagger_k\hat A_k$, we find
\begin{equation}
\hat A^\dagger_k\hat A_k=\frac{1}{2M}\{\hat p^2-i\hbar k[\hat p,\tan(k\hat x)]+\hbar^2 k^2\tan^2(k\hat x)\}.
\label{eq: adaga}
\end{equation}
Computing the commutator is usually done by choosing to work in the coordinate representation, where the momentum operator is proportional to a derivative with respect to $x$. It can also be evaluated algebraically (without resorting to any representation, or using calculus), which is shown in Appendix A; the motivation for this work is a wonderful early paper by Dirac~\cite{dirac_1926}, which develops the algebra for the radial momentum operator similar to how we do in Appendix B. Plugging in the result from Eq.~(\ref{eq: tan}) $[\hat p,\tan(k\hat x)]=-i\hbar k \sec^2(k\hat x)$, and recognizing that $\tan^2(k\hat x)+1=\sec^2(k\hat x)$, yields
\begin{equation}
\hat A^\dagger_k\hat A_k=\frac{\hat p^2}{2M}-\frac{\hbar^2 k^2}{2M}.
\label{eq: adaga2}
\end{equation}
This implies that the energy term $E_0$ satisfies $E_0=\hbar^2k^2/2M$ in order to produce the original Hamiltonian in Eq.~(\ref{eq: factorized}). Since we need to pick the largest $k$ value before divergence of the operator (to yield the largest $E_0$ value, according to the requirements of the factorization method), we pick $k=\pi/L$. This tells us the ground state, which satisfies
\begin{equation}
\hat A_{k=\pi/L}|\psi_{gs}\rangle=0,
\label{eq: ground-state}
\end{equation}
has energy $E_0$. This follows because the lowest-energy state of the Hamiltonian, written as in Eq.~(\ref{eq: factorized}), has energy $E_0$, because the operator $\hat A^\dagger_k\hat A_k$ is a positive semi-definite operator, which has a minimal expectation value of 0, when Eq.~(\ref{eq: ground-state}) is satisfied.

Our next step is to find the wavefunction. Once again, a standard treatment would multiply Eq.~(\ref{eq: ground-state}) by $\langle x|$ and represent the momentum operator as a derivative to find a first-order differential equation for the ground state wavefunction, which is easily solved. Here, we proceed without calculus, and use the fact that the state $\langle x|$ can be written as the translation of the state $\langle x{=}0|$ via
\begin{equation}
\langle x|=\langle x{=}0|e^{i\frac{x\hat p}{\hbar}},
\label{eq: trans}
\end{equation}
with $x$ a number in the exponent.
We use the Hadamard identity (see Appendix A for a derivation without calculus) to first show that
\begin{equation}
\left [ e^{i\frac{x\hat p}{\hbar}},\hat x\right ]=xe^{i\frac{x\hat p}{\hbar}}
\label{eq: trans2}
\end{equation}
and then
\begin{equation}
\langle x|\hat x=\langle x{=}0|e^{i\frac{x\hat p}{\hbar}}\hat x=\langle x{=}0|(\hat x+x)e^{i\frac{x\hat p}{\hbar}}
=\langle x{=}0|e^{i\frac{x\hat p}{\hbar}}x=\langle x|x,
\label{eq: trans3}
\end{equation}
which follows because $\langle x{=}0|\hat x=0$. This sequence of equalities verifies that $\langle x{=}0|\exp(ix\hat p/\hbar)
=\langle x|$ because when we operate on this state with $\hat x$ from the right, we obtain the number $x$ multiplying the original state; i.e., it is an eigenvalue-eigenvector relation. 

The wavefunction then is found from the following steps: first, we use the translation operator to determine the state $\langle x|$
\begin{equation}
\psi_{gs}(x)=\langle x|\psi_{gs}\rangle=\langle x{=}0|e^{i\frac{x\hat p}{\hbar}}|\psi_{gs}\rangle;
\end{equation}
second, we expand the exponential in its power series (which also can be derived without calculus, by using the binomial theorem and the property that $e^xe^y=e^{x+y}$, but takes us too far afield to show the details here)
\begin{equation}
\psi_{gs}(x)=\langle x{=}0|\sum_{n=0}^\infty\frac{1}{n!}\left (\frac{ix}{\hbar}\right )^n(\hat p)^n|\psi_{gs}\rangle;
\end{equation}
and  third, we move the numbers out of the matrix element
\begin{equation}
\psi_{gs}(x)=\sum_{n=0}^\infty\frac{1}{n!}\left (\frac{ix}{\hbar}\right )^n\langle x{=}0|(\hat p)^n|\psi_{gs}\rangle.
\end{equation}
For the fourth step,  we need to evaluate the matrix elements. The even powers are easy, because $|\psi_{gs}\rangle$ is an eigenstate under $\hat p^2$ with eigenvalue $\hbar^2\pi^2/L^2$. Odd powers can use this eigenvalue relation to remove all operators except for one power of $\hat p$. We determine the action of $\hat p$ on $|\psi_{gs}\rangle$ from Eq.~(\ref{eq: ground-state}), which yields
\begin{equation}
\hat p |\psi_{gs}\rangle=i\hbar k\tan(k\hat x)|\psi_{gs}\rangle.
\end{equation}
Since we evaluate this state against the $\langle x{=}0|$ bra, we find that it vanishes due to $\langle x{=}0|\tan(k\hat x)=\langle x{=}0|\tan(k\times 0)=0$. So we find the wavefunction becomes
\begin{equation}
\psi_{gs}(x)=\sum_{n=0}^\infty (-1)^n\frac{1}{(2n)!}\left (\frac{\pi x}{L}\right )^{2n}\langle x{=}0|\psi_{gs}\rangle
=\cos\left (\frac{\pi x}{L}\right )\langle x{=}0|\psi_{gs}\rangle.
\end{equation}
The matrix element $\langle x{=}0|\psi_{gs}\rangle$ is a number, which provides the normalization constant for the wavefunction. This number can only be determined with calculus, but it's precise value is not needed for any of the discussions given in this work, and is often not needed for calculations (since it cancels when evaluating expectation values).

We have completed the standard derivation of the ground state for the particle in a box using no calculus. This step is identical to the methodology of Schr\"odinger. Going forward to find the other eigenstates takes us away from the original  Schr\"odinger methodology. Our alternative approach is simple---we adjust $k$ to other values that are consistent with the conditions of the problem being solved---keeping in mind that we must choose the largest constant $E_0$ in the factorization of the Hamiltonian if there is any ambiguity. So, we increase $k$ from the ground-state solution found with $k=\pi/L$ until the raising and lowering operators diverge again at the boundary, namely when $k=3\pi/L$. We continue in this fashion and find that the correct $k$ values are
\begin{equation}
k=\frac{(2n+1)\pi}{L},
\label{eq: k}
\end{equation}
with the associated wavefunctions
\begin{equation}
\psi_{2n+1}(x)=\cos\left (\frac{(2n+1)\pi x}{L}\right )\langle x{=}0|\psi_{2n+1}\rangle,
\label{eq: psi_k}
\end{equation}
and energies
\begin{equation}
E_{2n+1}=\frac{\hbar^2(2n+1)^2\pi^2}{2ML^2}.
\label{eq: excited_energy}
\end{equation}
We can immediately verify that these are the even wavefunction solutions for the particle in a box, where we now use the integer $2n+1$ to label the different solutions. One might be concerned about whether there are any problems associated with the fact that the raising and lowering operators diverge at internal points inside the box, but it turns out that these divergences occur precisely where the wavefunction vanishes, which requires us to evaluate these (raising/lowering) operators acting on the wavefunctions with a proper limiting procedure; doing so produces a finite value since the node of the wavefunction cancels the divergence of the operator. This procedure makes the calculation of the wavefunctions in the modified Schr\"odinger operator method follow a similar approach to the standard differential equation approach, where the physical solutions to the differential equation are chosen, by selecting only those that also solve the appropriate boundary condition at the edge of the box.

But these are not all of the solutions of the particle in a box. They are just the even solutions. To find the odd solutions, we need to find another factorization of $\hat\mathcal{H}$. Fortunately, this is easy to achieve. The lowering operator needed for the odd solutions is
\begin{equation}
\hat B_k=\frac{1}{\sqrt{2m}}\left [ \hat p+i\hbar k \,{\rm cotan}(k\hat x)\right ].
\label{eq: b_op}
\end{equation}
A quick calculation using Eq.~(\ref{eq: cotan})  and the trigonometric identity ${\rm cotan}^2(k\hat x)+1={\rm cosec}^2(k\hat x)$ yields
\begin{equation}
\hat B_k^\dagger \hat B_k=\frac{\hat p^2}{2M}-\frac{\hbar^2 k^2}{2M},
\label{eq: eq: b_ham}
\end{equation}
which tells us that $\hat\mathcal{H}=\hat B^\dagger_k \hat B_k+\hbar^2 k^2/2M$. What is our rule for choosing $k$? One immediately sees that the operator diverges at $x=0$. It will also diverge at the edges of the box when $k=2n\pi/L$. This is the condition to maximize the constant term for each interval of $k$ where the operator next diverges. Obviously, the eigenstate $|\phi_{2n}\rangle$ satisfies $\hat B_{2n\pi/L}|\phi_{2n}\rangle=0$  and has the corresponding energy $E_{2n}=\hbar^2(2n)^2\pi^2/2ML^2$.

The wavefunction cannot be derived in the same fashion as we did for the even functions above because ${\rm cotan}(0)=\infty$, making a power-series expansion about $x=0$ problematic. Instead, we show that there is a simple relationship between the wavefunction we want to find $|\phi_{2n}\rangle$ and an auxiliary wavefunction $|\psi_{2n}\rangle$, which satisfies $\hat A_{2n}|\psi_{2n}\rangle=0$. Note that $|\psi_{2n}\rangle$ does not satisfy the proper boundary condition, but we have already derived that
\begin{equation}
\psi_{2n}(x)=\langle x|\psi_{2n}\rangle=\cos\left (\frac{2n\pi x}{L}\right )\langle x{=}0|\psi_{2n}\rangle,
\label{eq: psi_2n}
\end{equation}
since the derivation did not use the boundary condition.
Next, we show that 
\begin{equation}
|\phi_{2n}\rangle=\tan(k\hat x)|\psi_{2n}\rangle.
\label{eq: phi_psi}
\end{equation}
This follows by establishing two facts. First, we verify that $\hat B_k\tan(k\hat x)|\psi_k\rangle=0$ via a direct computation:
\begin{eqnarray}
\hat B_k\tan(k\hat x)|\psi_k\rangle&=&\frac{1}{\sqrt{2M}}\left [ \hat p +i\hbar k\,{\rm cotan}(k\hat x)\right ] \tan(k\hat x)|\psi_k\rangle\nonumber\\
&=&\frac{1}{\sqrt{2M}}\left [ [\hat p,\tan(k\hat x)]+\tan(k\hat x)\hat p+i\hbar k\right ]|\psi_k\rangle\nonumber\\
&=&\frac{1}{\sqrt{2M}}\left [ -i\hbar k\sec^2(k\hat x)+\tan(k\hat k)i\hbar k \tan(k\hat x)+i\hbar k\right ]|\psi_k\rangle\nonumber\\
&=&\frac{1}{\sqrt{2M}}i\hbar k\left [ \frac{-1+\sin^2(k\hat x)+\cos^2(k\hat x)}{\cos^2(k\hat x)}\right ]|\psi_k\rangle=0.
\label{eq: phi_deriv}
\end{eqnarray}
Hence, $\tan(k\hat x)|\psi_k\rangle$ satisfies the defining relation for $|\phi_k\rangle$, given by $\hat B_k|\phi_k\rangle=0$. This means the two functions differ by at most a multiplicative constant. So, second, we choose the constant to provide a normalized wavefunction. Hence, we have
\begin{equation}
\phi_{2n}(x)=\tan\left (\frac{2n\pi x}{L}\right )\cos\left (\frac{2n\pi x}{L}\right )\langle 0|\psi_{2n}\rangle=
\sin\left (\frac{2n\pi x}{L}\right )\langle 0|\psi_{2n}\rangle,
\label{eq: phi_odd}
\end{equation}
which  completes the derivation of the odd wavefunctions.

This derivation for the particle-in-a-box problem is much simpler than the Schr\"odinger derivation, which requires an infinite chain of operator relationships and appears in many textbooks (see, for example, \cite{green} or \cite{ohanian}). 

It turns out that one can use this same methodology to solve for the wavefunctions of a particle restricted to move on a circle of circumference $L$. Here, instead of having the operator diverge at the endpoints, we require the operator to be periodic, so that $\hat A_k(\hat x+L)=\hat A_k(\hat x)$ (and similarly for $\hat B_k$). Note that we {\it do not} assume that the wavefunction must be continuous and hence periodic, as is often done. Indeed, contrary to the statements in many textbooks, {\it there is no fundamental principle that requires the wavefunction to be continuous}. The proper requirements are that the probability density be continuous and that the probability current be continuous. Both can be satisfied for this problem with {\it either} periodic boundary conditions for the wave function {\it or} with antiperiodic boundary conditions. This is seen automatically with the operator method, because the condition for periodicity of the operator is that $\tan(k\hat x+kL)=\tan(k\hat x)$ and similarly for the ${\rm cotan}$. Both are satisfied by $k=n\pi/L$ (periodic boundary conditions) or $k=(n+1/2)\pi/L$ (antiperiodic boundary conditions). We immediately see the periodic or antiperiodic boundary conditions arising from the forms for the wavefunctions as being proportional to $\cos(kx)$ for $\hat A_k$ and $\sin(kx)$ for $\hat B_k$. Of course, the resolution of this issue is that only periodic solutions are consistent with orbital angular momentum~\cite{green,ballentine}. Note that the energy takes the same form as before, with $E_k=\hbar^2k^2/2M$, but with these different allowed choices for $k$ (including now $k=0$ for the even solutions). We will describe more completely why continuity of the raising and lowering operators is required to conserve the probability current when we discuss the case with piecewise continuous potentials below.

We find it interesting that the operator method requires us to confront this issue about the properties of the wavefunction and provides a nice introduction to these subtleties that are often glossed over in textbooks. One could also discuss other subtle issues, such as the facts that the momentum operator $\hat p$ is Hermitian but not self-adjoint for the particle-in-a-box problem and the position operator $\hat x$ is Hermitian but not self-adjoint for the particle-on-a-circle problem. We will not discuss these issues in detail, nor will we discuss related issues about uncertainty relations and how they are modified for operators that are not self-adjoint. But one could motivate these discussions in a classroom setting if desired to discuss such issues. In most cases these Hermitian versus self-adjoint discussions are rather advanced and technical and best left for more advanced courses~\cite{qm_for_math}.

\section{Particle in a finite square-well one-dimensional box}

We move on to describing the particle in a finite box. The potential now goes to zero far from the origin, so we have that the Hamiltonian satisfies $\hat\mathcal{H}=\hat p^2/2m+\hat V(\hat x)$ with
\begin{equation}
\hat V(\hat x)=\left \{\begin{array}{c}
~~~~0~~{\rm for~}|x|\ge \frac{L}{2}\\
-V_0~~{\rm for~}|x|<\frac{L}{2}.
\end{array}
\right .
\label{eq: leaky_pot_1d}
\end{equation}
Here, the wavefunction is nonzero almost everywhere, so the consistency condition on the raising and lowering operators is different. The solution has already been briefly described~\cite{brazil}. We note here that one cannot proceed in the original Schr\"odinger fashion because nothing changes inside the box, so the energy levels would come out the same as those for the particle in an infinite box. These are not the correct energy levels. Instead, we proceed as we discussed above: (i) we first create a factorization of the Hamiltonian that depends on a parameter; (ii) we adjust the parameter to solve the consistency condition; and (iii) we use all operators, factorizations and energies to determine the eigenfunctions and eigenvalues of the problem.

Inside the box, the presence of a nonzero potential makes some small changes to the problem: We still use the same $\hat A_k$ and $\hat B_k$ operators for the even and odd solutions, respectively, but the energy is shifted by $V_0$ to yield the two equations
\begin{equation}
\hat\mathcal{H}=\hat A_k^\dagger\hat A_k+E_k
\label{eq: a_ham2}
\end{equation}
and
\begin{equation}
\hat\mathcal{H}=\hat B_k^\dagger\hat B_k+E_k
\label{eq: b_ham2}
\end{equation}
with $E_k=\hbar^2 k^2/2M-V_0$ in both cases. The bound states also satisfy $E_k\le 0$, which we also assume holds, since we will solve only for the bound states here. Hence, inside the box, we have $\hat A_k=[\hat p-i\hbar k\tan(k\hat x)]/\sqrt{2M}$ and $\hat B_k=[\hat p +i\hbar k \,{\rm cotan}(k\hat x)]/\sqrt{2M}$.

Outside the box, we need to find a new factorization for the Hamiltonian. It turns out that this factorization is rather simple---it satisfies $\hat C_{\pm \kappa}=[\hat p \pm i\hbar\kappa]/\sqrt{2M}$. For either choice of the sign, we find
\begin{equation}
\hat C_{\pm \kappa}^\dagger C_{\pm \kappa}=\frac{\hat p^2}{2M}+\frac{\hbar^2\kappa^2}{2M},
\label{eq: c_def}
\end{equation}
because $\kappa$ is a number and so it commutes with $\hat p$. Here, we have $E_k=-\hbar^2\kappa^2/2M$ holds, with the same $E_k$ value found inside the box.

Before finishing the problem, we need to determine what the wavefunction is outside the box with this new factorization. As before, the state that has energy $E_k$ satisfies $\hat C_{\pm \kappa}|\psi_k\rangle=0$. Hence, outside the box, we find
\begin{equation}
\hat p|\psi_k\rangle=\mp i\hbar\kappa|\psi_k\rangle.
\label{eq: c_eig}
\end{equation}
To find the wavefunction, we take the overlap with $\langle x|=\langle x{=}0|\exp(-ix\hat p)/\hbar)$. Because the eigenstate $|\psi_k\rangle$ satisfies an eigenvector-like relation outside the box under the operator $\hat p$, we immediately find that
\begin{equation}
\psi_k(x)=\langle x|\psi_k\rangle\propto e^{\mp \kappa x}.
\label{eq: c_wf}
\end{equation}
In order for the wavefunction to be normalizable, we must choose the plus sign for the operator when $x<-L/2$ and the minus sign when $x>L/2$; this then produces exponentially decaying functions for large $|x|$.

\begin{figure}[htb]
\centerline{
\includegraphics[width=4.5in]{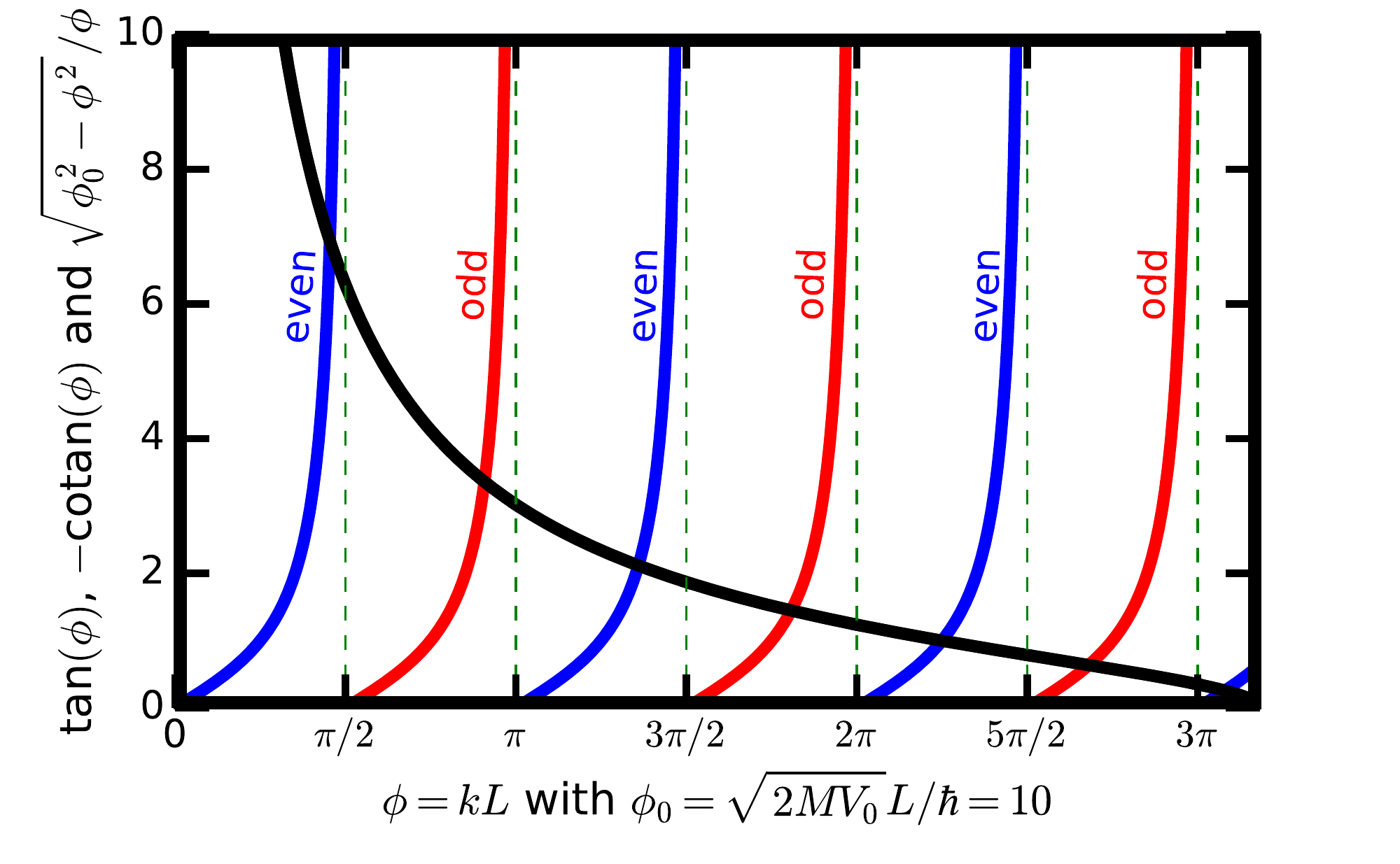}}
\caption{Graphical solution for the even and odd wavefunctions of the particle in a finite one-dimensional box. The blue lines are $\tan(\phi)$, the red lines are $-{\rm cotan}(\phi)$, and the black line is $\sqrt{\phi_0^2-\phi^2}/\phi_0$. The dashed lines indicate multiples of $\pi/2$. The parameter $\phi_0=10$ for this plot. \label{fig: trans_1d}}
\end{figure}

Finally, we require that the raising and lowering operators be continuous at $x=-L/2$ and at $x=L/2$; we will see below that this continuity condition guarantees the standard condition imposed via the Schr\"odinger equation approach, namely continuity of the logarithmic derivative of the wavefunction at the boundary. It may seem odd that one would require such a continuity condition for an operator that factorizes the Hamiltonian (which is discontinuous), but it is needed because it determines the wavefunction. We will see this condition always produces the proper results for quantum systems. It is related to the requirements of continuity of the probability density and the probability current, as follows: First, the probability density in one dimension is $|\langle x|\psi\rangle|^2$. For this to be continuous at a point $x_0$ where the potential is discontinuous requires
$\lim_{x\to x_0^+}|\langle x|\psi\rangle|^2=\lim_{x\to x_0^-}|\langle x|\psi\rangle|^2$. This is satisfied if $\lim_{x\to x_0^+}\langle x|\psi\rangle=\exp(i\alpha)\lim_{x\to x_0^-}\langle x|\psi\rangle$, with $\exp(i\alpha)$ a complex phase. Continuity of the probability current requires $\langle \psi|x\rangle\, \langle x|\hat p|\psi\rangle$ to be continuous as we approach $x=x_0$ from above or below. If we write $\hat A=[\hat p-i\hbar k W(\hat x)]/\sqrt{2m}$, with $W(\hat x)$ being the ``superpotential,'' then we see that continuity of the current occurs only if $\lim_{x\to x_0^-}W(x)=\lim_{x\to x_0^+}W(x)$, which is identical to saying that the lowering operator $\hat A$ is continuous at $x_0$. This is because the fact that $|\psi\rangle$ is annihilated by $\hat A$ means $\hat p|\psi\rangle\propto W(\hat x)|\psi\rangle$.  Of course, when we evaluate the wavefunctions below, we immediately see that they do satisfy the appropriate continuity conditions for both the probability and the probability current. 

This continuity condition is the same at each point and produces
\begin{equation}
	\tan\left (\frac{kL}{2}\right )=\frac{\kappa}{k}
\label{eq: tangent}
\end{equation}
for the even solutions and
\begin{equation}
-\,{\rm cotan}\left (\frac{kL}{2}\right )=\frac{\kappa}{k}
\label{eq: cotangent}
\end{equation}
for the odd solutions. These equations are solved in the standard fashion. We first define an angle $\phi$ via
\begin{equation}
\frac{kL}{2}=\phi=\frac{\sqrt{2M(V_0+E_k)}}{\hbar}\frac{L}{2}
\label{eq: phi_def}
\end{equation}
and a parameter $\phi_0$ by
\begin{equation}
\phi_0=\frac{\sqrt{2MV_0}}{\hbar}\frac{L}{2}.
\label{eq: phi0_def}
\end{equation}
Then, the two transcendental equations become
\begin{equation}
\tan\phi=\sqrt{\frac{\phi_0^2}{\phi^2}-1}
\label{eq: transcend1}
\end{equation}
and
\begin{equation}
-\,{\rm cotan}\,\phi=\sqrt{\frac{\phi_0^2}{\phi^2}-1}.
\label{eq: transcend2}
\end{equation}
Notice that the right hand side of both equations is the same. In Fig.~\ref{fig: trans_1d}, we plot the left and right hand sides of both equations. Points where they intersect correspond to solutions of the respective equations. 
 In Fig.~\ref{fig:  1d_energy}, we show the energy levels as a function of the dimensionless parameter that represents the potential, $\phi_0$.  The key element for this work is that regardless of the size of $\phi_0$, there is always at least one solution to the first transcendental equation in Eq.~(\ref{eq: transcend1}) because the tangent runs from $0$ to $\infty$ as $\phi$ runs from $0$ to $\pi/2$, implying it must intersect the value on the right hand side somewhere.

\begin{figure}[htb]
\centerline{
\includegraphics[width=4.5in]{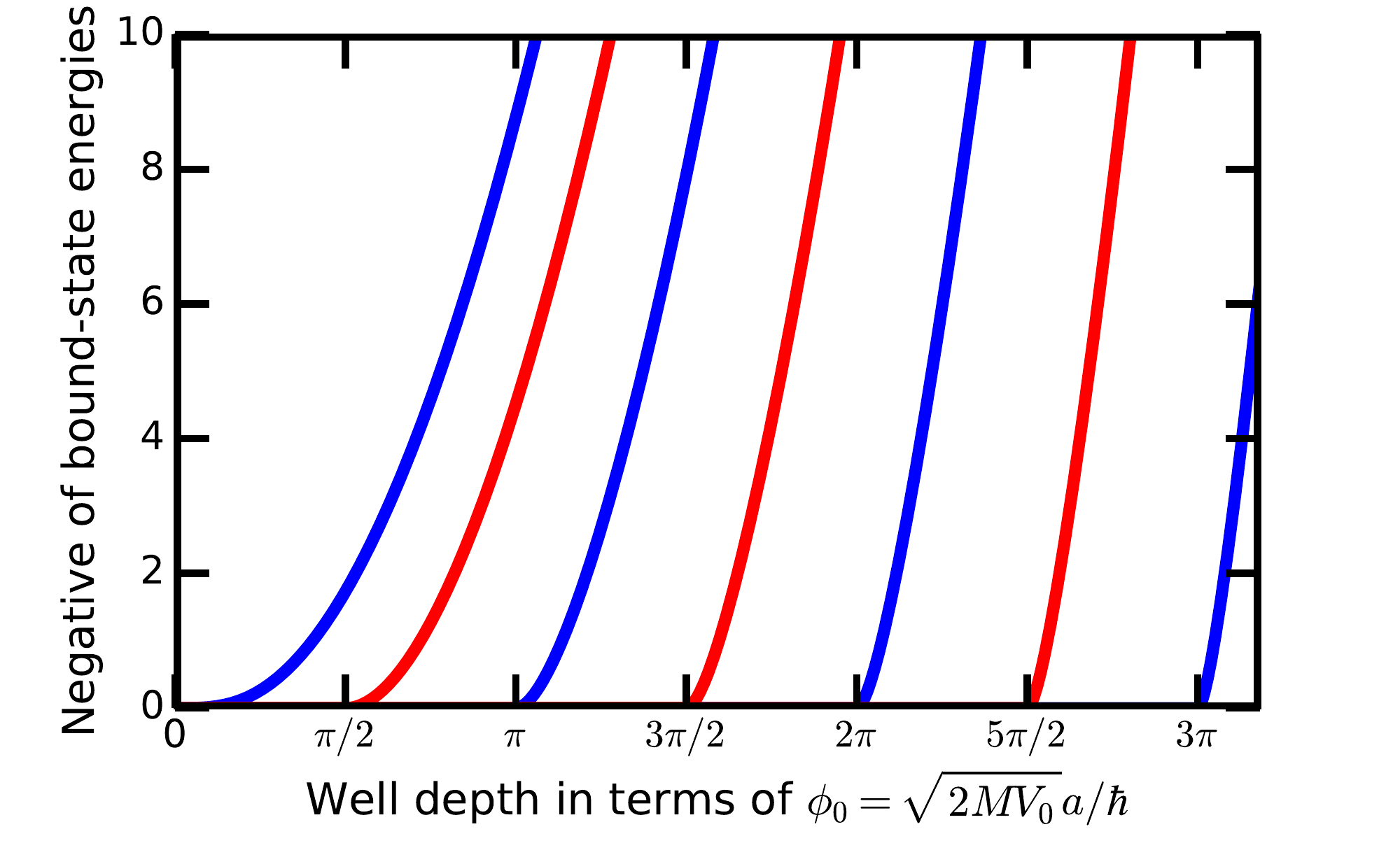}}
\caption{Energy eigenvalues for the particle in a one-dimensional finite square-well potential as a function of the potential well depth. A new bound state appears every time the parameter $\phi_0$ increases past a half integer multiple of $\pi$. Blue curves are the even solutions and red ones the odd, just like in Fig.~\ref{fig: trans_1d}. \label{fig: 1d_energy}}
\end{figure}

Summarizing, the Hamiltonian for the particle in a finite box can be written in one of two factorizations. Even solutions have
\begin{equation}
\hat A_k=\left \{
\begin{array}{l l}
\frac{1}{\sqrt{2M}}(\hat p +i\hbar \kappa)&{\rm for~}x<-\frac{L}{2}\\
\frac{1}{\sqrt{2M}}[\hat p-i\hbar k\tan(k\hat x)]&{\rm for~}|x|\le \frac{L}{2}\\
\frac{1}{\sqrt{2M}}(\hat p-i\hbar\kappa)&{\rm for~}x>\frac{L}{2}.
\end{array}
\right .
\label{eq: even_op_1d}
\end{equation}
For every $\phi^*$ which solves Eq.~(\ref{eq: transcend1}), the wavefunction becomes
\begin{equation}
\psi_k^e(x)=\left \{
\begin{array}{l l}
e^{\kappa x}\langle x{=}0|\phi_k\rangle&{\rm for~}x<-\frac{L}{2}\\
\cos(kx)\langle x{=}0|\phi_k\rangle&{\rm for~}|x|<\frac{L}{2}\\
e^{-\kappa x}\langle x{=}0|\phi_k\rangle&{\rm for~}x>\frac{L}{2}
\end{array}
\right .
\label{eq: wf_even_1d}
\end{equation}
with $k=2\phi^*/L$. 
The number $\langle x{=}0|\phi_k\rangle$ is the normalization constant (which we have not determined) and
the energy of the state is $\hbar^2k^2/2m-V_0$. Odd solutions have
\begin{equation}
\hat B_k=\left \{
\begin{array}{l l}
\frac{1}{\sqrt{2M}}(\hat p +i\hbar \kappa)&{\rm for~}x<-\frac{L}{2}\\
\frac{1}{\sqrt{2M}}[\hat p+i\hbar k\,{\rm cotan}(k\hat x)]&{\rm for~}|x|\le \frac{L}{2}\\
\frac{1}{\sqrt{2M}}(\hat p-i\hbar\kappa)&{\rm for~}x>\frac{L}{2}.
\end{array}
\right .
\label{eq: odd_op_1d}
\end{equation}
For each $\phi^*$ which solves Eq.~(\ref{eq: transcend2}), the wavefunction becomes
\begin{equation}
\psi_k^o(x)=\left \{
\begin{array}{l l}
-e^{\kappa x}\langle x{=}0|\phi_k\rangle&{\rm for~}x<-\frac{L}{2}\\
\sin(kx)\langle x{=}0|\phi_k\rangle&{\rm for~}|x|<\frac{L}{2}\\
e^{-\kappa x}\langle x{=}0|\phi_k\rangle&{\rm for~}x>\frac{L}{2}
\end{array}
\right .
\label{eq: wf_odd_1d}
\end{equation}
with $k=2\phi^*/L$. One can easily verify that the wavefunction and its derivative (or more naturally, the logarithmic derivative) is continuous at the point where the potential is discontinuous.

\section{Particle in a finite square-well two-dimensional circular box}

Having finished the one-dimensional case, we now move on to the two-dimensional case. Interestingly, the particle-in-a-finite-square-well problem is not separable for a rectangular box, but it is for a circular box, which is what we consider now.
The potential is
\begin{equation}
V(\hat  r,\hat \theta)=\left \{
\begin{array}{l}
-V_0~~{\rm for~}r\le R\\
~~~0~~~{\rm for~}r>R.
\end{array}
\right .
\end{equation}
Using the form for the kinetic energy in two-dimensions given in Eq.~(\ref{eq: kinetic3}), the Hamiltonian
is
\begin{equation}
\hat\mathcal{H}=\frac{\hat p_r^2}{2M}+\frac{\hat L_z^2}{2M\hat r^2}-\frac{\hbar^2}{8M\hat r^2}-V_0\,\theta(R-\hat r),
\label{eq: 2d_ham}
\end{equation}
where the theta function is the Heaviside unit step function and we use a capital $M$ for the mass, so as not to confuse with the $z$-component of angular momentum below. 

We will organize our states under the eigenvalue of the $z$-component of angular momentum, whose eigenvalue is $\hbar m$, with $m$ an integer. (This is a standard procedure in quantum mechanics, so we do not include the details here.) Our first step is to work on the solution for $m=0$, which has vanishing $z$-component of angular momentum.
We claim that the operator 
\begin{equation}
\hat A_k=\frac{1}{\sqrt{2M}}\left [ \hat p_r-i\hbar k \frac{J_1(k\hat r)}{J_0(k\hat r)}+i\frac{\hbar}{2\hat r}\right ]
\label{eq: ak_def_2d}
\end{equation}
is the lowering operator for the case where the eigenvalue under $\hat L_z$ is zero. To check, we need to evaluate the following:
\begin{equation}
\hat A_k^\dagger\hat A_k=\frac{1}{2M}\left \{ \hat p_r^2-i\hbar k \left [\hat p_r,\frac{J_1(k\hat r)}{J_0(k\hat r)}\right ]+i\hbar\left [\hat p_r,\frac{1}{2\hat r}\right ]+\left [\hbar k\frac{J_1(k\hat r)}{J_0(k\hat r)}-\frac{\hbar}{2\hat r}\right ]^2\right\}.
\label{eq: ak_2d_com}
\end{equation}
Here the $J_m$ functions are Bessel functions of the first kind, discussed in Appendix C.
The evaluation of the commutator needs to be done in steps. We use the product rule to find
\begin{equation}
\left [\hat p_r,\frac{J_1(k\hat r)}{J_0(k\hat r)}\right ]=J_1(k\hat r)\left [\hat p_r,\frac{1}{J_0(k\hat r)}\right ]+\left [\hat p_r,J_1(k\hat r)\right ]\frac{1}{J_0(k\hat r)},
\label{eq: com_bessel_j01}
\end{equation}
where the second commutator is evaluated with Eq.~(\ref{eq: bessel_identity1}). The first commutator is evaluated as we have evaluated similar ones before using a ``multiply by one'' trick:
\begin{eqnarray}
0&=&\left [\hat p_r,\frac{J_0(k\hat r)}{J_0(k\hat r)}\right ]=J_0(k\hat r)\left [\hat p_r,\frac{1}{J_0(k\hat r)}\right ]+
[\hat p_r,J_0(k\hat r)]\frac{1}{J_0(k\hat r)}\nonumber\\
0&=&J_0(k\hat r)\left [\hat p_r,\frac{1}{J_0(k\hat r)}\right ]+i\hbar k J_1(k\hat r)\frac{1}{J_0(k\hat r)},
\label{eq: com_invbessel}
\end{eqnarray}
so that
\begin{equation}
\left [\hat p_r,\frac{1}{J_0(k\hat r)}\right ]=-i\hbar k\frac{ J_1(k\hat r)}{[J_0(k\hat r)]^2}.
\label{eq: com_invbessel2}
\end{equation}

We now have everything needed to simplify the result in Eq.~(\ref{eq: ak_2d_com}):
\begin{eqnarray}
\hat A_k^\dagger\hat A_k &=&\frac{1}{2M}\left \{ \hat p_r^2-\hbar^2 k^2 \frac{[J_1(k\hat r)]^2}{[J_0(k\hat r)]^2}+\hbar^2 k^2\frac{J_2(k\hat r)-J_0(k\hat r)}{2J_0(k\hat r)}-\frac{\hbar^2}{2\hat r^2}\right .\nonumber\\
&~&~~~~~+\left .\left [\hbar k\frac{J_1(k\hat r)}{J_0(k\hat r)}-\frac{\hbar}{2\hat r}\right ]^2\right\}\nonumber\\
&=&\frac{1}{2M}\left \{ \hat p_r^2+\hbar^2 k^2\frac{\frac{1}{k\hat r}J_1(k\hat r)-J_0(k\hat r)}{J_0(k\hat r)}
-\frac{\hbar^2}{4\hat r^2}-\frac{\hbar^2 k}{\hat r}\frac{J_1(k\hat r)}{J_0(k\hat r)}\right \}\nonumber\\
&=&\frac{\hat p_r^2}{2M}-\frac{\hbar^2}{8M\hat r^2}-\frac{\hbar^2 k^2}{2M},
\end{eqnarray}
where in the second equality, we employed the identity in Eq.~(\ref{eq: bessel_identity2}) with $m=1$ to replace $J_2$
with $J_1$ and $J_0$, and in the last line, we simplified the result. If we define $E_k=\hbar^2 k^2/2M-V_0$, then we have $\hat\mathcal{H}=\hat A_k^\dagger\hat A_k+E_k$ for $\hat r\le R$, in the case where the $\hat L_z$ eigenvalue is zero. 

The derivation of the result for $\hat r>R$ is similar, but we need to use the modified Bessel functions of the second kind, $K_m$, which exponentially decay for large argument. Their properties are stated at the end of Appendix C. We define $\kappa$ via $E_k=-\hbar^2\kappa^2/2M$ (recall the energy is less than zero because it is a bound state). Then the lowering operator in this region becomes
\begin{equation}
\hat C_\kappa=\frac{1}{\sqrt{2M}}\left [ \hat p_r-i\hbar\kappa\frac{K_1(\kappa \hat r)}{K_0(\kappa\hat r)}+i\frac{\hbar}{2\hat r}\right ],
\label{eq: c_kappa}
\end{equation}
where we replaced $k$ by $\kappa$ and $J$ by $K$. Calculating the operator $\hat C^\dagger_\kappa\hat C_\kappa$
proceeds just as we did before and yields (with the identities in Appendix C)
\begin{equation}
\hat C^\dagger_\kappa \hat C_\kappa=\frac{\hat p_r^2}{2M}+\frac{\hbar^2}{8M\hat r^2}+\frac{\hbar^2\kappa^2}{2M}.
\label{eq: a_kappa_dag_a_kappa}
\end{equation}
This implies that $\hat\mathcal{H}=\hat C^\dagger_\kappa \hat C_\kappa+E_k$ for $\hat r>R$. 

\begin{figure}[htb]
\centerline{
\includegraphics[width=4.5in]{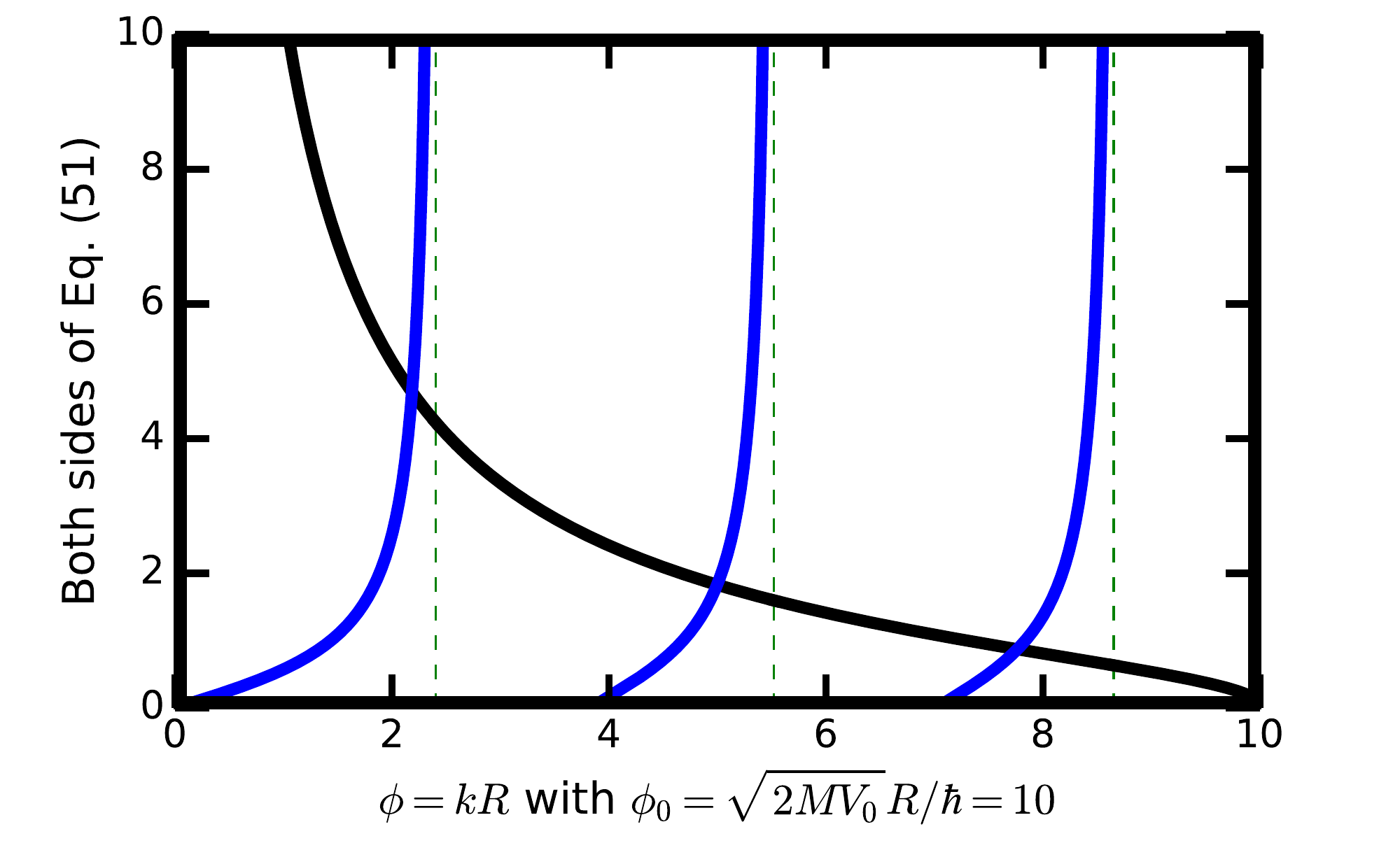}}
\caption{Graphical solution to the transcendental equation from Eq.~(\ref{eq: 2d_transcend2}) for the solutions to the particle in a finite square-well potential in two dimensions. The blue curves are the left hand side $J_1(\phi)/J_0(\phi)$, while the black curve is the right hand side $\sqrt{\phi_0^2-\phi^2}K_1(\sqrt{\phi_0^2-\phi^2})/[\phi K_0(\sqrt{\phi_0^2-\phi^2})]$. Points of intersection are the solutions of the equation. The dashed lines are the zeros of the Bessel function $J_0(\phi)$, where the left hand side diverges. \label{fig: transcend_2}}
\end{figure}

As before, we require the lowering operator to be continuous at $\hat r=R$, which yields
\begin{equation}
-i\hbar k\frac{J_1(kR)}{J_0(kR)}+i\frac{\hbar}{2R}=-i\hbar\kappa\frac{K_1(\kappa R)}{K_0(\kappa R)}+i\frac{\hbar}{2R},
\label{eq: 2d_transcend1}
\end{equation}
and simplifies to
\begin{equation}
\frac{J_1(kR)}{J_0(kR)}=\frac{\kappa}{k}\frac{K_1(\kappa R)}{K_0(\kappa R)}.
\label{eq: 2d_transcend2}
\end{equation}
The two sides of this equation are plotted in Fig.~\ref{fig: transcend_2}, so that the solution corresponds to the points where the curves cross. We use $kR=\phi$ and $\sqrt{2MV_0}R/\hbar=\phi_0$, similar to how we treated the one-dimensional case (note as well that $\kappa R=\sqrt{\phi_0^2-\phi^2}$). This equation always has at least one solution. We illustrate this explicitly for the case where $V_0$ is small. In this case, both $\phi$ and $\phi_0$ are also small. We use the asymptotic behavior of the Bessel functions for small argument to learn that $J_0(\phi)\approx 1$, $J_1(\phi)\approx\phi/2$, $K_0\left (\sqrt{\phi_0^2-\phi^2}\right )\approx -\ln\left (\sqrt{\phi_0^2-\phi^2}\right )$, and $K_1\left (\sqrt{\phi_0^2-\phi^2}\right )\approx 1/\sqrt{\phi_0^2-\phi^2})$. Using these results, we find the transcendental equation becomes
\begin{equation}
\frac{\phi}{2}\approx -\frac{\sqrt{\phi_0^2-\phi^2}}{\phi}\frac{1}{\sqrt{\phi_0^2-\phi^2}\ln\left ({\sqrt{\phi_0^2-\phi^2}}\right )},
\label{eq: 2d_transcend3}
\end{equation}
or, after simplifying
\begin{equation}
\ln\left ({\sqrt{\phi_0^2-\phi^2}}\right )\approx -\frac{2}{\phi^2}.
\label{eq: 2d_transcend4}
\end{equation}
Re-expressing in terms of the energy, we find
\begin{equation}
E\approx -\frac{\hbar^2}{2MR^2}e^{-\frac{\hbar^2}{MR^2V_0}},
\label{eq: 2d_energy_final}
\end{equation}
which is an exponentially small result for small $V_0$.

So, we have shown that if we find a $k$ and $\kappa$ that solve Eq.~(\ref{eq: 2d_transcend2}), then the state that satisfies
both
\begin{equation}
\left [ \hat p_r-i\hbar k \frac{J_1(k\hat r)}{J_0(k\hat r)}+i\frac{\hbar}{2\hat r}\right ]|\phi_{k,\kappa}\rangle=0,
\label{eq: 2d_phi_state1}
\end{equation}
for $r\le R$, and
\begin{equation}
\left [ \hat p_r-i\hbar\kappa\frac{K_1(\kappa \hat r)}{K_0(\kappa\hat r)}+i\frac{\hbar}{2\hat r}\right ]|\phi_{k,\kappa}\rangle=0
\label{eq: 2d_phi_state2}
\end{equation}
for $r>R$ is the eigenstate, with an energy given by $E=-V_0+\hbar^2 k^2/2M=-\hbar^2\kappa^2/2M$.

\begin{figure}[htb]
\centerline{
\includegraphics[width=4.5in]{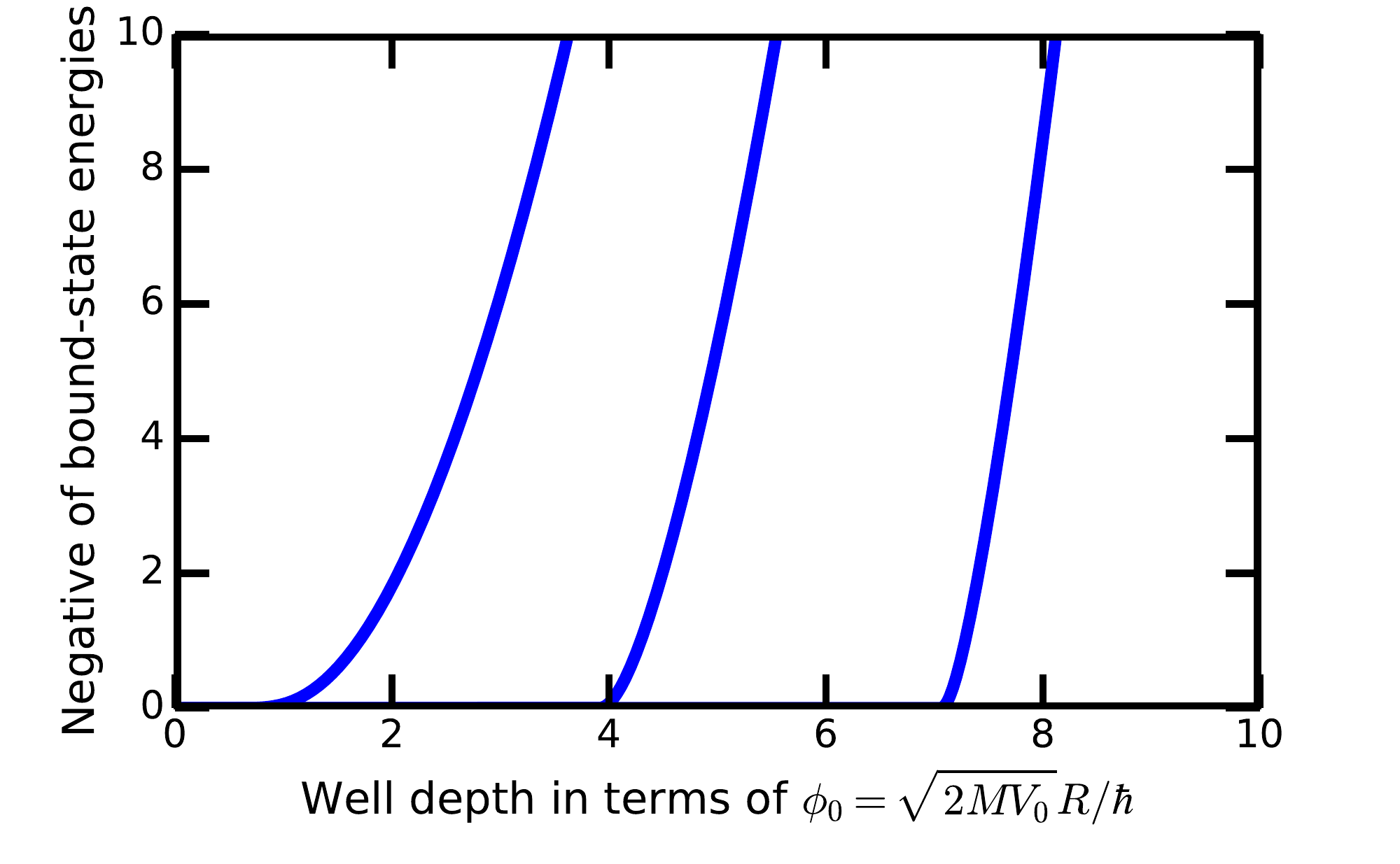}}
\caption{Energy eigenvalues for the particle in a two-dimensional circular finite square-well potential as a function of the potential well depth. A new bound state appears every time the parameter $\phi_0$ increases past a zero of the Bessel function where $J_1(\phi_0)=0$; the first three zeros are approximately at 3.832, 7.016, and 10.173. Initially, the bound state energy remains quite close to zero, and only emerges to larger values where one can see the curves ``take off'' in the figure. Blue curves plot the energies as a function of $\phi_0$. \label{fig: 2d_energy}}
\end{figure}

In Fig.~\ref{fig: 2d_energy}, we plot the corresponding (negative of the) energy levels as a function of $\phi_0$ for the two-dimensional case. One can see that the initial shape of the curve for small $\phi_0$ is quite flat, due to the exponentially small bound-state energy in this regime. Once the bound-state energy becomes sizable, the curves look quite similar to those of the one-dimensional case.

We still need to derive the wavefunction. This requires us to take the overlap of $|\phi_{k,\kappa}\rangle$ with $\langle r,m{=}0|$. Because the radial momentum operator is Hermitian, but not self-adjoint, one needs to proceed carefully to determine the translation operator for the radial coordinate. In Appendix D, we show how to do this, resulting in Eq.~(\ref{eq: r0_state}). Hence, we have
\begin{equation}
\phi_{k,\kappa}(r,\theta)=\langle r{=}0,\theta{=}0|e^{\frac{i}{\hbar}r\left (\hat p_r+i\frac{\hbar}{2\hat r}\right )}e^{\frac{i}{\hbar}\theta\hat L_z}|\phi_{k,\kappa}\rangle.
\label{eq: 2d_m0_wf1}
\end{equation}
But, because this state has $m=0$, we immediately find that $\exp[i\theta\hat L_z/\hbar]|\phi_{k,\kappa}\rangle=|\phi_{k,\kappa}\rangle$. Let us next assume that we have $r\le R$. Then we have
\begin{equation}
\phi_{k,\kappa}(r,\theta)=\sum_{n=0}^\infty\frac{1}{n!}\left (\frac{ir}{\hbar}\right )^n\langle r{=}0,\theta{=}0|
\left (\hat p_r+i\frac{\hbar}{2\hat r}\right )^n|\phi_{k,\kappa}\rangle.
\label{eq: 2d_m0_wf2}
\end{equation}
One can immediately verify that all odd powers of $n$ will vanish, because they are proportional to 
a linear combination of $J_m(k\hat r)$ terms in the numerator which all have $m\ne 0$; when evaluated against $\langle r{=}0|$ they all vanish. The even powers are not zero because their expansion includes terms with $J_0(k\hat r)$ which gives one when evaluated against the $\langle r{=}0|$ state. We need to determine the coefficient of the $J_0$ term for arbitrary $n$ to determine the wavefunction. This is done by a construction similar to Pascal's triangle.

To begin, we evaluate the powers in turn as follows: for $n=1$, we use Eq.~(\ref{eq: 2d_phi_state1}) to obtain
\begin{equation}
\left (\hat p_r+i\frac{\hbar}{2\hat r}\right )|\phi_{k,\kappa}\rangle=i\hbar k\frac{J_1(k\hat r)}{J_0(k\hat r)}
|\phi_{k,\kappa}\rangle.
\end{equation}
As we mentioned above, when we evaluate this with the bra at the origin, it vanishes, as will all odd powers.
Next is $n=2$, here, to evaluate the second power, we must evaluate $\hat p_r+i\hbar/2\hat r$ against the state above. This is done by first evaluating the commutator $\hat p_r$ with the fraction of Bessel functions and then
we have the operator acting against the $\phi_{k,\kappa}$ state. This yields
\begin{eqnarray}
\left (\hat p_r+i\frac{\hbar}{2\hat r}\right )^2|\phi_{k,\kappa}\rangle&=&i\hbar k\left (\hat p_r+i\frac{\hbar}{2\hat r}\right )\frac{J_1(k\hat r)}{J_0(k\hat r)}
|\phi_{k,\kappa}\rangle\\
&=&i\hbar k\left \{\left [ \hat p_r,\frac{J_1(k\hat r)}{J_0(k\hat r)}\right ]+i\hbar k\frac{J_1^2(k\hat r)}{J_0^2(k\hat r)}\right \}|\phi_{k,\kappa}\rangle\nonumber\\
&=&(i\hbar k)^2 \left \{ \frac{1}{2}\frac{J_2(k\hat r)-J_0(k\hat r)}{J_0(k\hat r)}-\frac{J_1^2(k\hat r)}{J_0^2(k\hat r)}+\frac{J_1^2(k\hat r)}{J_0^2(k\hat r)}\right \}|\phi_{k,\kappa}\rangle\nonumber
\end{eqnarray}
and we see that the only term that survives is the term involving the commutator of $\hat p_r$ with the numerator. This result holds for all $n$. As we take more and more commutators, we simply continue to use the rule derived in 
Eq.~(\ref{eq: bessel_identity1}) to evaluate them. This continues for all terms {\it except} the $J_0$ terms, since they are actually a constant when divided by $J_0$  and yield zero when we evaluate their commutator. When $n$ is odd, the numerator is a linear combination of odd index Bessel functions starting with $n$ and ending with one and when it is even, it is a linear combination of even powers starting with $n$ and ending with zero. When we evaluate against the bra at the origin, all terms vanish except for the coefficient of the $J_0$ term. We construct a Pascal-like triangle below
\begin{equation}
\begin{array}{l c c c c c c c c | c}
& & & & & & & & &{\rm Coefficient}\\
& & & & & & & & & {\rm of~} J_0(k\hat r)\\
m&8&7&6&5&4&3&2&1&0\\
n=1& & & & & & & & 1\\
n=2& & & & & & & \frac{1}{2}&  &-\frac{1}{2}\\
n=3& & & & & & \frac{1}{4}& &-\frac{3}{4}\\
n=4& & & & & \frac{1}{8}& &-\frac{4}{8}& &\frac{3}{8}\\
n=5& & & & \frac{1}{16}& &-\frac{5}{16}& &\frac{10}{16}& \\
n=6& & & \frac{1}{32}& &-\frac{6}{32}& &\frac{15}{32}& &-\frac{10}{32}\\
n=7& &\frac{1}{64}& &-\frac{7}{64}& &\frac{21}{64}& &-\frac{35}{64}\\
n=8&\frac{1}{128}& &-\frac{8}{128}& &\frac{28}{128}& &-\frac{56}{128}& &\frac{35}{128}
\end{array}
\end{equation}
For example, we worked out the case of $n=2$ which equals $[J_2(k\hat r)-J_0(k\hat r)]/2J_0(k\hat r)$ and the triangle has a 1/2 in the second entry $(m=2)$ and a -1/2 in the zeroth entry ($m=0$). The rules for constructing the ``triangle'' are as follows: The leftmost element of the row is equal to one-half the leftmost element in the row above it. Every other element in the row (except possibly the last one) is constructed by finding the two parents in the row above, subtracting the left parent from the right and dividing the result by two. If the new row is an odd row, we construct the rightmost entry by a different rule---it is equal to twice the right parent minus the left parent, with the result divided by two. If the new row is an even row, the rightmost entry is minus one half the rightmost entry of the row above (left parent). For example, the $n=6$ row has its second entry given by $(-5/16-1/16)/2=-6/32$ and so on. While the $n=7$ row has its rightmost entry given by $[2\times (-10/32)-15/32]/2=-35/64$. Note that the table satisfies the property that the sum of the absolute values of the entries in each row is equal to one.

We can actually find an explicit formula for the elements in each row according to this construction. For an odd row with $n=2i+1$, the nonzero $m=2j+1$ entry is
\begin{equation}
\left (\frac{1}{2}\right )^{2i}\frac{(2i+1)!}{(i+j+1)!(i-j)!}(-1)^{i-j}
\end{equation}
and for $n=2i$, the nonzero $m=2j$ entry is
\begin{equation}
\left (\frac{1}{2}\right )^{2i-1}\frac{(2i)!}{(i+j)!(i-j)!}(-1)^{i-j}
\end{equation}
for $j\ne 0$ and the $j=0$ entry is
\begin{equation}
\left (\frac{1}{2}\right )^{2i}\frac{(2i)!}{i!i!}(-1)^i.
\end{equation}
We can verify these results by induction, but do not go through those steps here. The coefficient of the $n=2i$ term in Eq.~(\ref{eq: 2d_m0_wf2}) becomes
\begin{equation}
\frac{1}{(2i)!}\left (\frac{ir}{\hbar}\right )^{2i}\left ( \frac{i\hbar k}{2}\right )^{2i}\frac{(2i)!}{i!i!}(-1)^i
=\left (\frac{kr}{2}\right )^{2i}\frac{1}{i!i!}(-1)^i,
\end{equation}
which is precisely what is needed to generate the $J_0(kr)$ Bessel function, so $\psi_{k,\kappa}(r)\propto J_0(kr)$ for $r<R$.

We need to go through a similar procedure for $r>R$ to generate the rest of the $m=0$ wavefunction. The difference is that we evaluate the terms in the series starting from the $\langle r{=}R|$ state. The result ends up being a Taylor series 
for the modified Bessel function about the point $r=R$. Unfortunately, there is no simple way to derive this without invoking some calculus (although for purists, it can be done by using properties of the Bessel functions, but this approach becomes quite tortuous). When we use the $k$ and $\kappa$ values that satisfy the transcendental equation, we verify that the wavefunction (determined from the factorization method) and its slope [or equivalently, the matrix element $\langle r,\theta|\hat p_r|\phi_{k,\kappa}\rangle$] are both continuous at $r=R$. This then agrees with the standard Schr\"odinger equation approach.

We next sketch how to solve the problem when the $z$-component of angular momentum is nonzero (we denote it by $\hbar m$). As shown in Appendix B, the general case has a Hamiltonian given by
\begin{equation}
\hat\mathcal{H}_m=\left \{\begin{array}{l l}
\frac{\hat p_r^2}{2M}+\frac{\hbar^2(m^2-\frac{1}{4})}{2M\hat r^2}-V_0 &{\rm for~}\hat r\le R\nonumber\\
\nonumber\\
\frac{\hat p_r^2}{2M}+\frac{\hbar^2(m^2-\frac{1}{4})}{2M\hat r^2} &{\rm for~}\hat r> R.
\end{array}
\right .
\label{eq: 2d_ham_general}
\end{equation}
We construct the solutions for nonzero $m$ by following the standard Schr\"odinger prescription. First, we define
a lowering operator via
\begin{equation}
\hat A_m=\frac{1}{\sqrt{2M}}\left [ \hat p_r+i\frac{\hbar\left (m-\frac{1}{2}\right )}{\hat r}\right ].
\label{eq: lowering_m}
\end{equation}
Then, we find the following two identities:
\begin{eqnarray}
\hat A^\dagger_m\hat A_m&=&\frac{\hat p_r^2}{2M}+i\frac{\hbar\left ( m-\frac{1}{2}\right )}{2M}\left [\hat p_r,\frac{1}{\hat r}\right ]+\frac{\hbar^2\left (m-\frac{1}{2}\right )^2}{2M\hat r^2}\nonumber\\
&=&\frac{\hat p_r^2}{2M}-\frac{\hbar^2\left ( m-\frac{1}{2}\right )}{2M\hat r^2}+\frac{\hbar^2\left (m-\frac{1}{2}\right )^2}{2M\hat r^2}\nonumber\\
&=&\frac{\hat p_r^2}{2M}+\frac{\hbar^2\left [(m-1)^2-\frac{1}{4}\right ]}{2M\hat r^2}
\label{eq: 2d_adagm_am}
\end{eqnarray}
and
\begin{eqnarray}
\hat A_m\hat A_m^\dagger&=&\frac{\hat p_r^2}{2M}-i\frac{\hbar\left ( m-\frac{1}{2}\right )}{2M}\left [\hat p_r,\frac{1}{\hat r}\right ]+\frac{\hbar^2\left (m-\frac{1}{2}\right )^2}{2M\hat r^2}\nonumber\\
&=&\frac{\hat p_r^2}{2M}+\frac{\hbar^2\left [m^2-\frac{1}{4}\right ]}{2M\hat r^2}.
\label{eq: 2d_am_adagm}
\end{eqnarray}
Using these identities, we have that $\hat\mathcal{H}_0=\hat A_1^\dagger\hat A_1-V_0\,\theta(R-\hat r)$ and $\hat A_m^\dagger \hat A_m=\hat A_{m-1}\hat A_{m-1}^\dagger$. We claim that if we take an eigenfunction $|\phi_{k,\kappa}\rangle$ which satisfies Eqs.~(\ref{eq: 2d_phi_state1}) and (\ref{eq: 2d_phi_state2}) [but we do not require it to solve the transcendental equation in Eq.~(\ref{eq: 2d_transcend2})], then $\hat A_m\hat A_{m-1}\cdots\hat A_2\hat A_1|\phi_{k,\kappa}\rangle$ is an eigenstate. We show this result directly. First, we write $\hat\mathcal{H}_m=\hat A_m\hat A_m^\dagger-V_0\,\theta(R-\hat r)$ and recognize that the potential, being one of two different constant values, commutes with all $\hat A_m$ operators (except possibly at one point, and for now we are working on verifying the wavefunction everywhere {\it except} at $r=R$). Then we first verify that the state
$|\psi_{m,k,\kappa}\rangle=\hat A_m\hat A_{m-1}\cdots\hat A_2\hat A_1|\phi_{k,\kappa}\rangle$ satisfies
\begin{eqnarray}
[\hat\mathcal{H}_m-V(\hat r)]|\psi_{m,k,\kappa}\rangle&=&(\hat A_m\hat A_m^\dagger)\hat A_m\hat A_{m-1}\cdots\hat A_2\hat A_1|\phi_{k,\kappa}\rangle\nonumber\\
&=&\hat A_m\hat (A_m^\dagger\hat A_m)\hat A_{m-1}\cdots\hat A_2\hat A_1|\phi_{k,\kappa}\rangle\nonumber\\
&=&\hat A_m(\hat A_{m-1}\hat A_{m-1}^\dagger)\hat A_{m-1}\cdots\hat A_2\hat A_1|\phi_{k,\kappa}\rangle\nonumber\\
&=&\hat A_m\hat A_{m-1}\hat A_{m-1}\cdots\hat A_2\hat A_1(\hat A_1^\dagger\hat A_1)|\phi_{k,\kappa}\rangle\nonumber\\
&=&\hat A_m\hat A_{m-1}\hat A_{m-1}\cdots\hat A_2\hat A_1[\hat\mathcal{H}_0-V(\hat r)]|\phi_{k,\kappa}\rangle,
\end{eqnarray}
where we used the identities in Eqs.~(\ref{eq: 2d_adagm_am}) and (\ref{eq: 2d_am_adagm}) to move the daggered operator as far as we could to the right. Next, we add the term with $V(\hat r)$ back in, noting it commutes with the $\hat A_m$ operators and find that
\begin{eqnarray}
\hat\mathcal{H}_m|\psi_{m,k,\kappa}\rangle&=&\left (\frac{\hbar^2k^2}{2M}-V_0\theta(R-\hat r)\right )|\psi_{m,k,\kappa}\rangle= -\frac{\hbar^2\kappa^2}{2M}|\psi_{m,k,\kappa}\rangle\nonumber\\
&=&E_{m,k,\kappa}|\psi_{m,k,\kappa}\rangle.
\end{eqnarray}
Hence, the state $|\psi_{m,k,\kappa}\rangle$ is an energy eigenstate of the Hamiltonian $\hat \mathcal{H}_m$. This procedure of moving the daggered operator through the undaggered chain is called an intertwining relation.

We still need to find a transcendental equation to determine $k$. We do this by requiring that the wavefunction and the matrix element of $\hat p_r$ are both continuous at $r=R$. Continuity of the matrix element of $\hat p_r$ is required for probability current conservation; as we saw above, we can choose the overall phase to be zero, and hence the wavefunction will also be continuous. While the procedure is completely straightforward for arbitrary $m$, we will illustrate how to do it only for the easiest cases given by $m=1$ and 2. The general case  follows by induction.

The case $m=1$ says
\begin{equation}
\psi_{m{=}1,k,\kappa}(r,\theta)=\langle r,\theta|\hat A_1\{ |\phi_{k,\kappa}\rangle\otimes|m{=}1\rangle\},
\end{equation}
where $|m{=}1\rangle$ is the eigenstate of $\hat L_z$ with eigenvalue $\hbar$. Substituting in for the operators and states, we immediately find that for $r\le R$ we have
\begin{equation}
\psi_{m{=}1,k,\kappa}(r,\theta)=\langle r|\frac{1}{\sqrt{2M}}\left (\hat p_r+i\frac{\hbar}{2\hat r}\right )|\phi_{k,\kappa}\rangle e^{i\theta}\langle \theta{=}0|m{=}1\rangle.
\end{equation}
Then Eq.~(\ref{eq: 2d_m0_wf1}) yields
\begin{equation}
\psi_{m{=}1,k,\kappa}(r,\theta)=\langle r|i\frac{\hbar k}{\sqrt{2M}}\frac{J_1(k\hat r)}{J_0(k\hat r)}|\phi_{k,\kappa}\rangle e^{i\theta}\langle \theta{=}0|m{=}1\rangle.
\end{equation}
Using the fact that $\langle r|\phi_{k,\kappa}\rangle\propto J_0(kr)$, we immediately find that $\psi_{m{=}1,k,\kappa}(r,\theta)\propto J_1(kr)\exp(i\theta)$.

The case for $m=2$ also proceeds similarly. Note that
\begin{equation}
\hat A_2\hat A_1|\phi_{k,\kappa}\rangle=i\frac{1}{\sqrt{2M}}\left (\hat p_r+i\frac{3\hbar}{2\hat r}\right )\frac{\hbar k}{\sqrt{2M}}\frac{J_1(k\hat r)}{J_0(k\hat r)}|\phi_{k,\kappa}\rangle.
\end{equation}
Commuting the $\hat A_2$ operator to the right to act on the state $|\phi_{k,\kappa}\rangle$ gives us
\begin{equation}
\hat A_2\hat A_1|\phi_{k,\kappa}\rangle=-\frac{\hbar^2k^2}{2M}\left ( \frac{J_2(k\hat r)-J_0(k\hat r)}{2J_0(k\hat r)}+\frac{1}{k\hat r}\frac{J_1(k\hat r)}{J_0(k\hat r)}\right )|\phi_{k,\kappa}\rangle.
\end{equation}
Next, we use the identity in Eq.~(\ref{eq: bessel_identity2}) to replace $J_1(k\hat r)/(k\hat r)$ by $J_0$ and $J_2$.
This results in
\begin{equation}
\hat A_2\hat A_1|\phi_{k,\kappa}\rangle=-\frac{\hbar^2k^2}{2M} \frac{J_2(k\hat r)}{J_0(k\hat r)}|\phi_{k,\kappa}\rangle.
\end{equation}
Hence, we have
\begin{equation}
\psi_{m{=}2,k,\kappa}(r,\theta)\propto J_2(kr)e^{2i\theta}
\end{equation}
for $r\le R$.

The general proof for arbitrary $m$ follows straightforwardly by induction and is not given in detail here.

We also need to work out the results for $r>R$. One can follow the strategy outlined above with the $J$ functions and find it also works for the $K$ functions (there are a few sign changes, but the end results are the same), and hence we have 
\begin{equation}
\psi_{m,k,\kappa}(r,\theta)\propto \left \{
\begin{array}{l}
J_m(kr)e^{im\theta}~~~{\rm for~}r\le R\\
K_m(kr)e^{im\theta}~~{\rm for~}r>R
\end{array}
\right .
\end{equation}
Continuity of the wavefunction at $r=R$ tells us that $J_m(kR)=K_m(\kappa R)$. Evaluating $\langle r|\hat p_r|\psi\rangle$ and ensuring it is continuous at $r=R$ requires 
\begin{eqnarray}
&~&\langle r| [\hat p_r,J_m(k\hat r)]\frac{1}{J_0(k\hat r)}-i\hbar \frac{J_m(k\hat r)}{2\hat r J_0(k\hat r)}|\phi\rangle=\nonumber\\
&~&~~~~~~~~~~~~~~~~~~~
\langle r|[\hat p_r,K_m(\kappa\hat r)]\frac{1}{K_0(\kappa \hat r)}-i\hbar\frac{K_m(\kappa\hat r)}{2\hat r K_0(\kappa \hat r)}|\phi\rangle. 
\end{eqnarray}
This tells us we must also have $kJ_{m+1}(kR)=\kappa K_{m+1}(\kappa R)$. Dividing the two equations yields the final result, which determines $k$ and $\kappa$
\begin{equation}
\frac{J_{m+1}(kR)}{J_m(kR)}=\frac{\kappa}{k}\frac{K_{m+1}(\kappa R)}{K_m(\kappa R)}.
\label{eq: 2d_final_transcend}
\end{equation}
The equation agrees with the previous transcendental equation we derived for $m=0$ in Eq.~(\ref{eq: 2d_transcend2}).

Using this last transcendental equation, we now check to see under what circumstances the system supports a bound state for nonzero $m$. We rewrite the equation in terms of $\phi$ and $\phi_0$ as
\begin{equation}
\frac{J_{m+1}(\phi)}{J_m(\phi)}=\frac{\sqrt{\phi_0^2-\phi^2}}{\phi}\frac{K_{m+1}\left (\sqrt{\phi_0^2-\phi^2}\right )}{K_m\left (\sqrt{\phi_o^2-\phi^2}\right )}.
\end{equation}
We use the facts that $J_m(x)\approx (x/2)^{m}/m!$ for small $x$ and $K_m(x)\approx (m-1)!(2/x)^{m}/2$ for small $x$ with $m\ge 1$ to analyze the behavior for small $V_0$. This tells us the solution exists if
\begin{equation}
\frac{\phi}{2(m+1)}=\frac{\sqrt{\phi_0^2-\phi^2}}{\phi}\frac{2m}{\sqrt{\phi_0^2-\phi^2}}
\end{equation}
which says $\phi^2=4m(m+1)$. But $\phi$ is small and $m\ge 1$, so there is no solution for small arguments. Hence, $V_0$ must be large enough for a solution to exist. We do not investigate the question of how big analytically.

\section{Particle in a finite square-well three-dimensional spherical box} 
We now move onto the final problem that we tackle with the operator formalism, namely finding the bound states for a particle in a finite three-dimensional box. The original solution to this problem was worked out by Peierls in 1929, but used a different methodology~\cite{peierls}. We use the operator method here and find it is quite similar to the particle in a one-dimensional box {\it and} the higher $m$ solutions we worked with in the  two-dimensional case.

We begin with the Hamiltonian. While one can derive the kinetic energy without calculus similar to what we did in two-dimensions, the techniques are closely related, so we do not reproduce them here. We simply state the results that the Hamiltonian for a particle in a finite square-well three-dimensional box is
\begin{equation}
\hat \mathcal{H}=\left \{
\begin{array}{l}
\frac{\hat p_r^2}{2M}+\frac{\hat{\vec{L}}\cdot\hat{\vec{L}}}{2M\hat r^2}-V_0~~{\rm for~}\hat r\le R\\
 \\
\frac{\hat p_r^2}{2M}+\frac{\hat{\vec{L}}\cdot\hat{\vec{L}}}{2M\hat r^2}~~~~~~~~~{\rm for~}\hat r> R,
\end{array}
\right . 
\label{eq: 3d_ham}
\end{equation}
where $\hat{\vec{L}}=\hat{\vec{r}}\times\hat{\vec{p}}$ is the angular momentum operator.
If we evaluate the Hamiltonian on states of definite total angular momentum, then the eigenvalue of the square of the angular momentum operator is $\hbar^2\, l(l+1)$, so the Hamiltonian becomes
\begin{equation}
\hat \mathcal{H}_l=\left \{
\begin{array}{l}
\frac{\hat p_r^2}{2M}+\frac{\hbar^2\, l(l+1)}{2M\hat r^2}-V_0~~{\rm for~}\hat r\le R\\
 \\
\frac{\hat p_r^2}{2M}+\frac{\hbar^2\, l(l+1)}{2M\hat r^2}~~~~~~~~~{\rm for~}\hat r> R.
\end{array}
\right . 
\label{eq: 3d_ham2}
\end{equation}

We first solve the case with $l=0$. Note that the Hamiltonian (as a function of the radial coordinate only) is simply that of a particle in a one-dimensional finite box, with one exception---the radius is never less than zero. So there is a hard wall boundary at $r=0$, where the wavefunction must vanish. If we look back at the derivation for the particle in a box, the odd solutions automatically vanish at $r=0$, so those are the solutions here. We quickly recap how the solutions work. The operators are the same, but the wavefunctions are different, because the radial coordinate, the radial momentum, and the translation operator for the radial coordinate, are different from the one-dimensional coordinate, momentum and translation operators. 

The $l=0$ lowering operator for $r\le R$ is then
\begin{equation}
\hat B=\frac{1}{\sqrt{2M}}[\hat p_r+i\hbar k\, {\rm cotan}(k\hat r)],
\label{eq: 3d_l0_op}
\end{equation}
which satisfies
\begin{equation}
\hat\mathcal{H}_{l=0}=\hat B^\dagger\hat B+\frac{\hbar^2 k^2}{2M}-V_0.
\label{eq: 3d_ham3}
\end{equation}
Similarly, the lowering operator for $r>R$ is
\begin{equation}
\hat C=\frac{1}{\sqrt{2M}}(\hat p_r-i\hbar \kappa),
\label{eq: 3d_l0_op2}
\end{equation}
which satisfies
\begin{equation}
\hat H_{l=0}=\hat C^\dagger\hat C+\frac{\hbar^2\kappa^2}{2M}.
\label{eq: 3d_ham4}
\end{equation}
We require the operators to be continuous at $\hat r=R$, which yields
\begin{equation}
k\,{\rm cotan}(kR)=\kappa.
\end{equation}
Using $\phi=kR$ and $\phi_0=\sqrt{2MV_0}R/\hbar$, the transcendental equation becomes
\begin{equation}
-{\rm cotan}\phi=\frac{\sqrt{\phi_0^2-\phi^2}}{\phi},
\end{equation}
which has a solution only when $\phi_0\ge \pi/2$, or $V_0\ge \hbar^2\pi^2/(8MR^2)$. Hence, the three-dimensional particle in a finite box requires a minimal attractive potential before it supports a bound state. 

We calculate the wavefunction in a similar fashion to what was done before. There is no angular dependence, and for the radial dependence, we need to use the correct translation operator (which can be derived similar to what we did for the two-dimensional case). Hence,
\begin{equation}
\langle r|=\langle r{=}0|e^{\frac{i}{\hbar}r\left (\hat p_r+i\frac{\hbar}{\hat r}\right )}
\end{equation}
and
\begin{eqnarray}
\phi_{l{=}0}(r)&=&\langle r|\phi_{l{=}0}\rangle=\langle r{=}0|e^{\frac{i}{\hbar}r\left (\hat p_r+i\frac{\hbar}{\hat r}\right )}|\phi_{l{=}0}\rangle\nonumber\\
&=&\sum_{n=0}^\infty\frac{1}{n!}\left (\frac{ir}{\hbar}\right )^n\langle r{=}0|\left (\hat p_r+i\frac{\hbar}{\hat r}\right )^n|\phi_{l{=}0}\rangle.
\label{eq: 3d_l0_wf1}
\end{eqnarray}
We begin with $r\le R$ to evaluate the wavefunction. We determine how $\hat p_r$ acts on the state $|\phi_{l{=}0}\rangle$ by using the fact that $\hat B|\phi_{l{=}0}\rangle=0$ to yield
\begin{equation}
\hat p_r|\phi_{l{=}0}\rangle=-i\hbar k\, {\rm cotan}(k\hat r)|\phi_{l{=}0}\rangle.
\label{eq: 3d_pr_on_phi}
\end{equation}
Similar to the one-dimensional case, we can show that all odd powers vanish and all even powers give
\begin{equation}
\langle r{=}0|\left (\hat p_r+i\frac{\hbar}{\hat r}\right )^{2n}|\phi_{l{=}0}\rangle=\frac{1}{2n+1}(\hbar k)^{2n}\langle r{=}0|\phi_{l{=}0}\rangle.
\end{equation}
The details are technical and are shown in Appendix E. Substituting into Eq.~(\ref{eq: 3d_l0_wf1}), we find for $r\le R$
\begin{equation}
\phi_{l{=}0}(r)=\sum_{n=0}^\infty(-1)^n\frac{1}{(2n+1)!}(kr)^{2n}\langle r{=}0|\phi_{l{=}0}\rangle=\frac{\sin(kr)}{kr}\langle r{=}0|\phi_{l{=}0}\rangle.
\label{eq: 3d_l0_wf2}
\end{equation}
Next, we move on to $r>R$. Here, the wavefunction satisfies $\hat p_r|\phi_{l{=}0}\rangle=i\hbar\kappa|\phi_{l{=}0}\rangle$. 
In this case, we need to modify Eq.~(\ref{eq: 3d_l0_wf1}) to perform the expansion about $r=R$:
\begin{eqnarray}
\phi_{l{=}0}(r)&=&\langle r|\phi_{l{=}0}\rangle=\langle r{=}R|e^{\frac{i}{\hbar}(r-R)\left (\hat p_r+i\frac{\hbar}{\hat r}\right )}|\phi_{l{=}0}\rangle\nonumber\\
&=&\sum_{n=0}^\infty\frac{1}{n!}\left (\frac{i(r-R)}{\hbar}\right )^n\langle r{=}R|\left (\hat p_r+i\frac{\hbar}{\hat r}\right )^n|\phi_{l{=}0}\rangle.
\label{eq: 3d_l0_wf3}
\end{eqnarray}
The result for the $n$th power of $(\hat p_r+i\hbar/\hat r)$ acting on the state is
\begin{equation}
\left (\hat p_r+i\frac{\hbar}{\hat r}\right )^n|\phi_{l{=}0}\rangle=(i\hbar\kappa)^n\sum_{m=0}^n\frac{n!}{(n-m)!}\frac{1}{(\kappa\hat r)^m}|\phi_{l{=}0}\rangle,
\end{equation}
for $r>R$.
The derivation is also rather technical and is given in Appendix E. Using this result, we conclude that
\begin{eqnarray}
\phi_{l{=}0}(r)&=&\phi_{l{=}0}(R)\sum_{n=0}^\infty\frac{1}{n!}[-\kappa(r-R)]^n\sum_{m=0}^n\frac{n!}{(n-m)!}\frac{1}{(\kappa R)^m}\nonumber\\
&=&\phi_{l{=}0}(R))\sum_{n=0}^\infty\sum_{m=0}^n\frac{[-\kappa (r-R)]^{n-m}}{(n-m)!}\frac{[-\kappa(r-R)]^m}{(\kappa R)^m}\nonumber\\
&=&\phi_{l{=}0}(R)\sum_{m=0}^\infty\sum_{n=m}^\infty\frac{[-\kappa (r-R)]^{n-m}}{(n-m)!}\frac{[-\kappa(r-R)]^m}{(\kappa R)^m}\nonumber\\
&=&\phi_{l{=}0}(R)\sum_{m=0}^\infty\sum_{n=0}^\infty\frac{[-\kappa (r-R)]^{n}}{n!}\frac{[-\kappa(r-R)]^m}{(\kappa R)^m}\nonumber\\
&=&\phi_{l{=}0}(R)e^{-\kappa(r-R)}\sum_{m=0}^\infty (-1)^m\left (\frac{r-R}{R}\right )^m,
\end{eqnarray}
where we simplified and regrouped terms in the second line, switched the order of the summation in the third line, let $n\rightarrow n+m$ in the fourth line, and summed the series over $n$ in the fifth line.
If we assume $R<r<2R$, then we can sum the remaining series to find
\begin{equation}
\phi_{l{=}0}(r)=\phi_{l{=}0}(R)e^{-\kappa(r-R)}\frac{1}{1+\frac{r-R}{R}}=\phi_{l{=}0}(R)e^{-\kappa(r-R)}\frac{\kappa R}{\kappa r}\propto \frac{e^{-\kappa r}}{\kappa r}.
\end{equation}
Since the summed series is an analytic function, we can ``analytically continue'' this result and verify that it holds for all $r>R$.

Hence, we have established that 
\begin{equation}
\phi_{l{=}0}(r)=\left \{
\begin{array}{l}
\alpha\frac{\sin(kr)}{kr}~~~~{\rm for~}r\le R\\
 \\
\beta\frac{e^{-\kappa r}}{\kappa r}~~~~~~{\rm for~}r>R.
\end{array}
\right .
\end{equation}
The ratio of $\alpha/\beta$ is determined by the solution of the transcendental equation, which relates $k$ and $\kappa$. The overall constant is then determined by normalization, a step we omit here.

The results for higher $l$ all have equal or  higher energies than the $l=0$ state because the contribution from the centrifugal barrier is positive semi-definite. Hence, we need not work out their solutions, even though they follow in a similar fashion to how we found the nonzero $m$ eigenvalues and eigenfunctions for the two-dimensional case. The key point is that we have found a potential in three-dimensions that does not have a bound state unless the attractive potential has a large enough magnitude.

\section{Variational proof of the existence of at least one bound state in one and two dimensions}

Most of the pedagogical literature on using the variational approach to prove that attractive potentials in one and two dimensions always have at least one bound state work from a variational approach with a specific trial wavefunction~\cite{kocher,lange_dellano,perez,brownstein,buell_shadwick,borghi}. Usually this wavefunction has a simple peak with an adjustable width (although it can be more complicated for the two-dimensional case). This approach is also employed in textbooks, as described in the introduction. 

The variational principle that we employ is straightforward and requires no calculus to derive. We assume that the orthonormal eigenstates are written as $|n\rangle$ with ${\mathcal H}|n\rangle=E_n|n\rangle$, with $E_n$ ordered such that $E_0\le E_i\le E_2\cdots$. Then, we examine a normalized state $|\psi\rangle=\sum_{n=0}^\infty c_n|n\rangle$ with $\sum_{n=0}^\infty |c_n|^2=1$. We compute directly that
\begin{equation}
\langle\psi |{\mathcal H}|\psi\rangle=\sum_{n=0}^\infty |c_n|^2E_n\ge \sum_{n=0}^\infty |c_n|^2E_0=E_0.
\end{equation}
So, we learn that the expectation value of the Hamiltonian on any trial wavefunction produces an energy greater than or equal to the ground-state energy. Turned around, the ground-state energy is less than the expectation values of the Hamiltonian for any normalized trial wavefunction $|\psi\rangle$.

\begin{figure}[htb]
\centerline{
\includegraphics[width=4.5in]{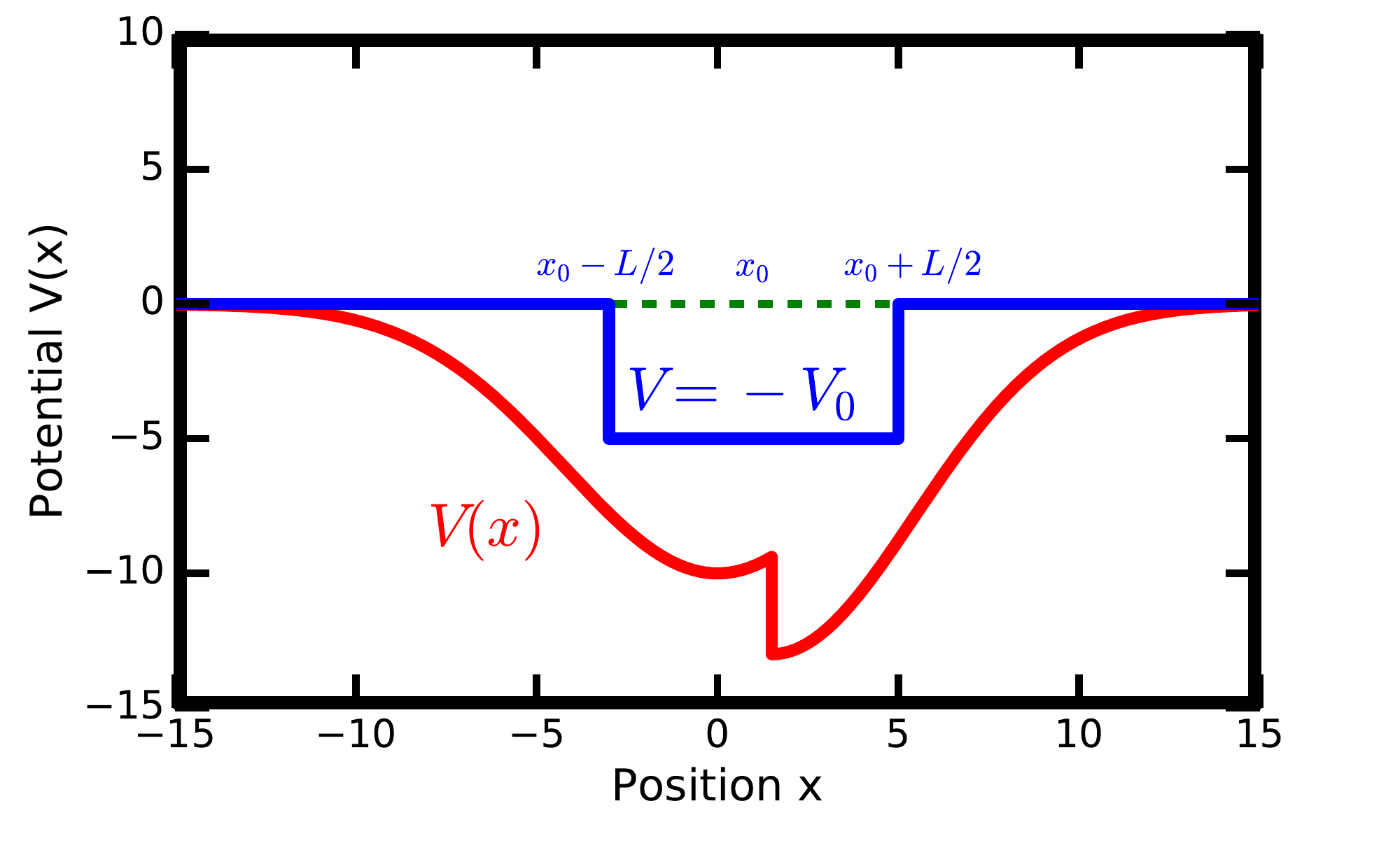}
}
\caption{Schematic picture of the set-up for the variational proof. The arbitrary attractive potential is depicted in red. It needs only to be nonpositive, to go to zero as $|x|\rightarrow \infty$ and to be piecewise continuous. The finite square-well box is depicted in blue. We choose an $x_0$, $L$, and $V_0$, so that the square-well potential ``fits inside'' the target potential, in the sense that $V_{\rm box}(x)\ge V(x)$ pointwise. The square well vanishes outside the box. \label{fig: variational}
}
\end{figure}

Next, we use the existence of a ground state for the particle-in-a-box problems to then prove that the ground state exists for arbitrary potentials which can have a box potential fit inside of them, as illustrated schematically in Fig.~\ref{fig: variational}. To be more precise, mathematically, we will prove potentials $V$ which satisfy the following criteria, all have at least one bound state. The criterion in one dimension is that
\begin{equation}
V(x)\le V_{\rm box}(x;V_0,L,x_0)\le 0,
\label{eq: 1d_criterion}
\end{equation}
where $V_{\rm box}(x;V_0,L,x_0)$ is the square-well potential of width $L$, depth $V_0$ and centered at the point $x=x_0$. Note that the description above is using the potential expressed as a function of the coordinate, not as an operator, so that the inequality is well-defined. In addition, we have the freedom to adjust the depth, width, and center in order to have it fit inside the target potential.  We allow the target potential to be piecewise continuous, to have divergences, etc., as long as it satisfies the criterion above. Similarly, for the two-dimensional potential, we require
\begin{equation}
V(x,y)\le V_{\rm box}(x,y;V_0,R,x_0,y_0)\le 0.
\label{eq: 2d_criterion}
\end{equation}
Here, $V_{\rm box}(x,y;V_0,R,x_0,y_0)$ is a circular ``square-well'' potential of depth $V_0$, radius $R$, and centered at $(x_0,y_0)$. 
Note in particular that this requirement implies that the arbitrary potential is always attractive. It is well known that this requirement is not needed for a bound state, and a weaker criterion that the integral of the potential over all space is less than zero is all that is needed~\cite{brownstein}. But that proof requires working with adjustable variational wavefunctions and is beyond the methodology we present here, which is algebraic and uses no calculus. So we work with the stricter requirement here.

Note that because the potential goes to zero as we approach infinity, a bound state requires $E<0$. This is because when $E<0$, the eigenstate will exponentially decay in the forbidden regions, as we have seen in the explicit solutions we worked out for the particle in finite square-well potentials. Our strategy is to prove that the ground state energy must be less than zero for systems with arbitrary attractive potentials that satisfy the required criteria. This then implies that those systems have at least one bound state.

The proof now follows very simply. For one dimension, we find a square-well potential that fits into the potential we want to show has a bound state according to Eq.~(\ref{eq: 1d_criterion}). Then, we compute the expectation value of the Hamiltonian for the arbitrary potential with the ground-state wavefunction of the box, which we denote as $|\phi_{\rm box}\rangle$. We find (using $\hat\mathcal{H}=\hat T+V(\hat x)$)
\begin{eqnarray}
\langle \phi_{\rm box}|\hat\mathcal{H}|\phi_{\rm box}\rangle&=&\langle \phi_{\rm box}|\hat T|\phi_{\rm box}\rangle+\langle\phi_{\rm box}|V(\hat x)|\phi_{\rm box}\rangle\nonumber\\
&\le& \langle \phi_{\rm box}|\hat T|\phi_{\rm box}\rangle+\langle\phi_{\rm box}|V_{\rm box}(\hat x,V_0,L,x_0)|\phi_{\rm box}\rangle\nonumber\\
&=&E_{\rm gs,box}(V_0,L)<0.
\end{eqnarray}
The basic idea is that the kinetic energy expectation value is the same for both the target Hamiltonian and the box Hamiltonian (when evaluated in the box ground state), and the expectation value of the potential is lower due to the criterion above. So the expectation value of the total energy is less than zero. Since all unbound states have an expectation value of the energy that is larger than zero (because their energy can always be written as $\hbar^2k^2/2M$ for some $k$), this proves the existence of at least one bound state.

The proof for the two-dimensional case is essentially the same, but we compare to the particle in a circular ``square-well.'' Of course, the proof does not hold for three dimensions because we found a counter-example with the particle in a spherical box.

To work on the more complex problem of asking how many bound states exist for a given potential, one must use more complex approaches that count the nodes of the wavefunction~\cite{ttwu} and are beyond the approach we have developed here which uses no calculus.

\section{Conclusions}

The goal of this work was to illustrate how one can solve particle-in-finite-square-well problems using operator methods with no calculus. We then applied those solutions to immediately prove that in one and two dimensions any fully attractive potential always has at least one bound state. The methodology we employed requires only high-school level math. It does require one to be able to work with power-series expansions, but only in an algebraic fashion. We also were able to demonstrate that the Schr\"odinger operator method can be streamlined for these problems. Schr\"odinger himself called his procedure ``shooting sparrows with artillery''~\cite{schrodinger_op2}. And his solution is indeed quite complex. The version we developed here, and hinted at in Refs.~\cite{brazil} and \cite{iran}, is much simpler and is quite similar to the conventional differential equation approach.

Some may complain that the derivation of the wavefunction itself still involves ``shooting sparrows with artillery.'' While we do not completely disagree with that characterization, using the same methods, we can derive the solutions in one, two, and three dimensions. The two-dimensional case is often not covered in textbooks, because it involves Bessel functions, which are viewed by many as a more advanced topic. Here, the procedure is unified, and the derivations share similar complexity across the different dimensions.

In the end, we feel that illustrating how far one can go with quantum-mechanical reasoning without needing to invoke advanced mathematics, can aid in bringing quantum phenomena to more people and help as well to show the important quantum principles without being buried in complex mathematics. Of course, if this restriction is relaxed, then one can derive the wavefuntions quite efficiently using only first-order differential equations. We did not discuss how to do this here because it already appears in many textbooks~\cite{green,ohanian,binney_skinner}.

We hope that this work will encourage others to discover additional areas in the quantum curriculum that do not require calculus. This will then provide choice to students as to how they would like to best learn the material. For example, we have already shown how one can determine spherical harmonics with only algebraic methods~\cite{weitzman}. Instead of thinking of operator methods as an advanced topic, we encourage others to think of it as a great way to {\it introduce and explore} quantum ideologies.

\section*{Acknowledgments}

We thank Wes Mathews for a critical reading of this manuscript.
J.~K.~F.~was supported by the National Science Foundation under grant number PHY-1620555 and by the McDevitt bequest at Georgetown University. Maurice Curran worked on the survey of seventy quantum textbooks. The remaining authors worked on all aspects of the paper.

\appendix

\section{Computing commutators without calculus}

In this appendix, we show how to compute the commutators $[\hat p,\tan(k\hat x)]$ and $[\hat p,{\rm cotan}(k\hat x)]$ without using calculus. It requires a few steps. First we use the Hadamard identity for the similarity transformation of the momentum operator
\begin{equation}
e^{-ik\hat x}\hat p e^{ik\hat x}=\hat p-ik[\hat x,\hat p]-\frac{k^2}{2}[\hat x,[\hat x,\hat p]]+\cdots+
\frac{(-ik)^n}{n!}[\hat x,[\hat x, \cdots,[\hat x,\hat p]\cdots]]_n\cdots,
\label{eq: hadamard}
\end{equation}
where the nth term has an n-fold nested commutator. In this case, the expansion terminates after two terms because the commutator of $\hat x$ with $\hat p$ is a number, which subsequently commutes with everything. We thereby find
\begin{equation}
e^{-ik\hat x}\hat p e^{ik\hat x}=\hat p+\hbar k,
\label{eq: p_sim}
\end{equation}
or 
\begin{equation}
[e^{-ik\hat x},\hat p]=\hbar k e^{-ik\hat x}.
\label{eq: exp_com}
\end{equation}
Next, we use the Euler relation to write $\exp(-ik\hat x)=\cos(k\hat x)-i\sin(k\hat x)$ and take the real and imaginary parts of both sides of Eq.~(\ref{eq: exp_com}) to find
\begin{equation}
[\cos(k\hat x),\hat p]=-i\hbar k\sin(k\hat x)
\label{eq: cos}
\end{equation}
and
\begin{equation}
[\sin(k\hat x),\hat p],=i\hbar k\cos(k\hat x).
\label{eq: sin}
\end{equation}
Note that this derivation uses the fact that the commutator of a real function of $\hat x$ with $\hat p$ is imaginary. Note further that these results agree with the standard commutation results we would find by using $\hat p=-i\hbar\partial_x$. 
The next step uses the product rule for a commutator and the ``multiply by one'' trick:
\begin{equation}
0=[1,\hat p]=\left [ \frac{\sin(k\hat x)}{\sin(k\hat x)},\hat p\right ]=\sin(k\hat x)\left [\frac{1}{\sin(k\hat x)},\hat p\right ]+[\sin(k\hat x),\hat p]\frac{1}{\sin(k\hat x)}
\label{eq: inv_sin}
\end{equation}
and then re-arranges the result to show that
\begin{equation}
\left [\frac{1}{\sin(k\hat x)},\hat p\right ]=-\frac{1}{\sin(k\hat x)}[\sin(k\hat x),\hat p]\frac{1}{\sin(k\hat x)}=-i\hbar k\,{\rm cosec}(k\hat x){\rm cotan}(k\hat x).
\label{eq: inv_sin2}
\end{equation}
Similarly, one can immediately verify
\begin{equation}
\left [\frac{1}{\cos(k\hat x)},\hat p\right ]=i\hbar k\,{\rm sec}(k\hat x){\rm tan}(k\hat x).
\label{eq: inv_cos2}
\end{equation}
The commutator for the tangent follows from Eqs.~(\ref{eq: sin}) and (\ref{eq: inv_cos2}) and the product rule for the commutator via
\begin{eqnarray}
[\tan(k\hat x),\hat p]&=&\frac{1}{\cos(k\hat x)}[\sin(k\hat x),\hat p]+\left [\frac{1}{\cos(k\hat x)},\hat p\right ]\sin(k\hat x)\nonumber\\
&=&i\hbar k+i\hbar k\tan^2(k\hat x)=i\hbar k\sec^2(k\hat x).
\label{eq: tan}
\end{eqnarray}
In the same fashion, we derive the commutator of the cotangent using Eqs.~(\ref{eq: cos}) and (\ref{eq: inv_sin2}) and the product rule. It is
\begin{eqnarray}
[{\rm cotan}(k\hat x),\hat p]&=&\frac{1}{\sin(k\hat x)}[\cos(k\hat x),\hat p]+\left [\frac{1}{\sin(k\hat x)},\hat p\right ]\cos(k\hat x)
\nonumber\\
&=&-i\hbar k-i\hbar k\,{\rm cotan}^2(k\hat x)=-i\hbar k\,{\rm cosec}^2(k\hat x).
\label{eq: cotan}
\end{eqnarray}

\section{Radial momentum and the Laplacian}

In this appendix, we derive the radial momentum and then the kinetic energy operator in two dimensions. We illustrate how this is done without calculus. We will work with polar coordinates and show how one converts from Cartesian to polar coordinates with quantum operators. The motivation for this appendix is one of Dirac's lesser known 1926 quantum mechanics publications~\cite{dirac_1926}, which shows how to employ Cartesian commutation relations to derive radial and polar relations.  

We begin with the square of the radial operator, defined by $\hat r^2=\hat r_x^2+\hat r_y^2$. From this, we can compute the commutator of the Cartesian momentum with $\hat r$. Let $\alpha$ denote $x$ or $y$. Then, (there is no Einstein summation convention used here)
\begin{equation}
[\hat r^2,\hat p_\alpha]=[\hat r_x^2+\hat r_y^2,\hat p_\alpha]=[\hat r_\alpha^2,\hat p_\alpha]=\hat r_\alpha[\hat r_\alpha,\hat p_\alpha]+[\hat r_\alpha,\hat p_\alpha]\hat r_\alpha=2i\hbar\hat r_\alpha.
\label{eq: rp_com1}
\end{equation}
Similarly, we have
\begin{equation}
[\hat r^2,\hat p_\alpha]=\hat r[\hat r,\hat p_\alpha]+[\hat r,\hat p_\alpha]\hat r.
\label{eq: rp_com2}
\end{equation}
It is well known that the commutator of a function of $\hat r_\alpha$ with $\hat p_\alpha$  is a function of $\hat r_\alpha$ only---it has no $\hat p_\alpha$ dependence. This follows by first noting that one can use induction to establish that $[(\hat r_\alpha)^n,\hat p_\alpha]=i\hbar n(\hat r_\alpha)^{n-1}$. Then, for any function that can be represented as a power series, we can employ this result to show that the commutator with $\hat p_\alpha$ is just a function of the coordinate operator. This implies that $[\hat r,[\hat r,\hat p_\alpha]]=0$, so we can move the $\hat r$ factor on the left of the first term on the right hand side of Eq.~(\ref{eq: rp_com2}) to the right. Then we use Eq.~(\ref{eq: rp_com1}) to immediately learn that
\begin{equation}
[\hat r,\hat p_\alpha]=i\hbar\frac{\hat r_\alpha}{\hat r}.
\label{eq: rp_com3}
\end{equation}
Using the product rule and the ``multiply by one'' trick, we also find that
\begin{equation}
[1,\hat p_\alpha]=0=\left [\frac{\hat r}{\hat r},\hat p_\alpha\right ]=\hat r\left [\frac{1}{\hat r},\hat p_\alpha\right ]+[\hat r,\hat p_\alpha]\frac{1}{\hat r}.
\label{eq: rp_com4}
\end{equation}
Combining with Eq.~(\ref{eq: rp_com3}) then shows that
\begin{equation}
\left [ \frac{1}{\hat r},\hat p_\alpha\right ]=-i\hbar\frac{\hat r_\alpha}{\hat r^3}.
\label{eq: rp_com5}
\end{equation}

We are now ready to determine the radial momentum. In classical mechanics, we simply take the dot product of the momentum vector with a unit vector in the radial direction. In quantum mechanics, we need to worry about operator ordering, so we find the proper way to find a Hermitian radial momentum operator is to average the two different orderings, so the radial momentum becomes
\begin{equation} 
\hat p_r=\frac{1}{2}\left ( \frac{\hat{\vec{r}}}{\hat r}\cdot \hat{\vec{p}}+\hat{\vec{p}}\cdot \frac{\hat{\vec{r}}}{\hat r}\right )= \frac{\hat{\vec{r}}}{\hat r}\cdot \hat{\vec{p}}-i\hbar\frac{1}{2\hat r}=\frac{1}{\hat r}\left (\hat x \hat p_x+\hat y\hat p_y-\frac{i\hbar}{2}\right ).
\label{eq: rad_mom}
\end{equation}
This radial momentum operator is canonically conjugate to $\hat r$. We compute
\begin{equation}
[\hat r,\hat p_r]=\left [ \hat r, \frac{1}{\hat r}\left (\hat x \hat p_x+\hat y\hat p_y\right )\right ]=\frac{i\hbar}{\hat r}\left (\hat x\frac{\hat x}{\hat r}+\hat y\frac{\hat y}{\hat r}\right )=i\hbar.
\label{eq: r_canon}
\end{equation}
Note that this component of the momentum operator does not depend solely on momentum, so a momentum eigenstate is {\it not} an eigenstate of the radial momentum.

We also need to find the $\theta$ component of the momentum. The unit vector perpendicular to $\vec{e_r}=\hat{\vec{r}}/\hat r$ is 
\begin{equation}
\vec{e_\theta}=-\frac{\hat y}{\hat r}\vec{e_x}+\frac{\hat x}{\hat r}\vec{e_y}.
\label{eq: e_theta}
\end{equation}
In this case, there are no ordering issues, so we immediately find that
\begin{equation}
\hat p_\theta=\hat{\vec{p}}\cdot\vec{e_\theta}=\frac{1}{\hat r}(-\hat y\hat p_x+\hat x\hat p_y).
\label{eq: p_theta}
\end{equation}
A straightforward calculation shows that $[\hat r,\hat p_\theta]=0$. The theta-component of momentum is related to the $z$-component of angular momentum via $\hat L_z=\hat r\hat p_\theta$. Note that we have $[\hat p_r,\hat L_z]=0$ and that this implies that $[\hat p_r,\hat p_\theta]\ne 0$. So it will be more convenient in many cases to work with $\hat L_z$ instead of $\hat p_\theta$.

We have already determined three of the four variables we need for polar coordinates. Our fourth is the angle $\hat\theta$. Usually, this angle is defined via an inverse tangent, but we find the arc-cosine is better for our purposes. So, we define
\begin{equation}
\hat\theta=\frac{\hat y}{|\hat y|}\cos^{-1}\left (\frac{\hat x}{\hat r}\right ),
\label{eq: theta_def}
\end{equation}
where $\hat y/|\hat y|={\rm sgn}\,{\hat y}$ is the sign of $\hat y$. Note that we work with $|\hat y|$ by computing first with its square just as we did with $\hat r$. One immediately finds that $[|\hat y|,\hat p_y]=i\hbar\hat y/|\hat y|$ and hence $[{\rm sgn}\,\hat y,\hat p_y]=0$, which is well-defined. Using the definition of $\hat\theta$, we find that
\begin{equation}
e^{i\hat\theta}=\cos\hat\theta+i\sin\hat\theta=\frac{\hat x}{\hat r}+i\frac{\hat y}{\hat r}.
\label{eq: exp_itheta}
\end{equation}
Use this to compute $[\exp(i\hat\theta),\hat p_r]$ as follows:
\begin{equation}
[e^{i\hat\theta},\hat p_r]=\left [\frac{\hat x}{\hat r}+i\frac{\hat y}{\hat r},\hat p_r\right ]=\left [\frac{\hat x}{\hat r}+i\frac{\hat y}{\hat r},\frac{1}{\hat r}(\hat x\hat p_x+\hat y\hat p_y)\right ]
=0.
\label{eq: exp_itheta2}
\end{equation}
Rearranging the commutator and using the Hadamard identity, then yields
\begin{equation}
e^{i\hat\theta}\hat p_r e^{-i\hat\theta}=\hat p_r=\hat p_r+i[\hat\theta,\hat p_r]+\frac{(i)^2}{2!}[\hat\theta,[\hat\theta,\hat p_r]]+\cdots
\label{eq: exp_itheta3}
\end{equation}
This equality only holds if we have $[\hat\theta,\hat p_r]=0$.

There is one commutator remaining, $[\hat \theta,\hat p_\theta]$, but we prefer to work with $\hat L_z=\hat r\hat p_\theta$. We first compute
\begin{eqnarray}
&\,&e^{-i\hat\theta}\hat L_ze^{i\hat\theta}=\left (\frac{\hat x}{\hat r}-i\frac{\hat y}{\hat r}\right )\hat L_z\left (\frac{\hat x}{\hat r}+i\frac{\hat y}{\hat r}\right )=\frac{1}{\hat r^2}[\hat x\hat L_z\hat x+\hat y\hat L_z\hat y+i(\hat x\hat L_z\hat y-\hat y\hat L_z\hat x)]\nonumber\\
~~&\,&=\frac{1}{\hat r^2}\{(\hat x^2+\hat y^2)\hat L_z+\hat x[\hat L_z,\hat x]+\hat y[\hat L_z,\hat y]
+i\hat x[\hat L_z,\hat y]-i\hat y[\hat L_z,\hat x]\}.
\label{eq: lz_com}
\end{eqnarray}
Using $[\hat L_z,\hat x]=i\hbar\hat y$ and $[\hat L_z,\hat y]=-i\hbar \hat x$, we find
\begin{equation}
e^{-i\hat\theta}\hat L_ze^{i\hat\theta}=\frac{1}{\hat r^2}[\hat r^2\hat L_z+i\hbar (\hat x\hat y-\hat y\hat x)+\hbar\hat r^2]=\hat L_z+\hbar.
\label{eq: lz_com2}
\end{equation}
Now, using the Hadamard identity, we find that $[\hat \theta,\hat L_z]=i\hbar$, or $[\hat \theta,\hat p_\theta]=i\hbar/\hat r$, which is not a canonical commutation relation, showing again that $\hat L_z$ is the momentum canonically conjugate to $\hat \theta$ and the better operator to work with.

We are now ready to calculate the kinetic energy. Since the radial momentum and the theta component of the momentum are perpendicular in classical mechanics, we expect the quantum kinetic energy to be the sum of the squares of both operators, noting that there may be some quantum corrections due to ordering issues. First, we note that the kinetic energy satisfies
\begin{equation}
\hat T=\frac{1}{2m}(\hat p_x^2+\hat p_y^2).
\label{eq: kinetic}
\end{equation}
Next, we compute $\hat p_r^2$:
\begin{equation}
\hat p_r^2=\frac{1}{\hat r}\left (\hat x \hat p_x+\hat y\hat p_y-i\frac{\hbar}{2}\right )\frac{1}{\hat r}\left (\hat x \hat p_x+\hat y\hat p_y-i\frac{\hbar}{2}\right ).
\label{eq: prsquared}
\end{equation}
Our first step is to move the middle $1/\hat r$ to the left:
\begin{equation}
\hat p_r^2=\frac{1}{\hat r^2}\left (\hat x \hat p_x+\hat y\hat p_y+i\frac{\hbar}{2}\right )\left (\hat x \hat p_x+\hat y\hat p_y-i\frac{\hbar}{2}\right ).
\label{eq: prsquared2}
\end{equation}
Expanding gives
\begin{eqnarray}
\hat p_r^2&=&\frac{1}{\hat r^2}\left (\hat x \hat p_x\hat x\hat p_x+2\hat x\hat y\hat p_x\hat p_y+\hat y\hat p_y\hat y\hat p_y +\frac{\hbar^2}{4}\right )\nonumber\\
&=&\frac{1}{\hat r^2}\left (\hat x^2 \hat p_x^2+2\hat x\hat y\hat p_x\hat p_y+\hat y^2\hat p_y^2 +\frac{\hbar^2}{4}\right ).
\label{eq: prsquared3}
\end{eqnarray}
Next, we compute $\hat p_\theta^2$ recalling that $\hat p_\theta$ commutes with $\hat r$:
\begin{equation}
\hat p_\theta^2=\frac{1}{\hat r^2}(-\hat y\hat p_x+\hat x\hat p_y)(-\hat y\hat p_x+\hat x\hat p_y)
=\frac{1}{\hat r^2}(\hat y^2\hat p_x^2-2\hat x\hat y\hat p_x\hat p_y+\hat x^2\hat p_y^2).
\label{eq: pthetasquared}
\end{equation}
So, we find that
\begin{equation}
\hat p_r^2+\hat p_\theta^2=\hat p_x^2+\hat p_y^2+\frac{\hbar^2}{4\hat r^2},
\label{eq: kinetic2}
\end{equation}
which becomes
\begin{equation}
\hat T=\frac{\hat p_r^2}{2m}+\frac{\hat L_z^2}{2m\hat r^2}-\frac{\hbar^2}{8m\hat r^2}.
\label{eq: kinetic3}
\end{equation}
Unlike the three-dimensional case, there is a correction term to the kinetic energy due to operator ordering issues.

\section{Working with Bessel functions without calculus}

In this appendix, we derive a few of the important identities involving Bessel functions. The Bessel function of the first kind is defined from its power series via
\begin{equation}
J_m(x)=\left ( \frac{x}{2}\right )^m \sum_{n=0}^\infty (-1)^n \left (\frac{x}{2}\right )^{2n}\frac{1}{n!(n+m)!},
\label{eq: bessel_def}
\end{equation}
for all integers $m\ge 0$; while we define $J_{-|m|}(x)=(-1)^{|m|} J_{|m|}(x)$ for negative $m$. We can derive these functions by projecting a plane wave in two dimensions onto the basis given by $r$ and the $z$-component of angular momentum, but we do not go into those details here.

We need to derive some identities with respect to these Bessel functions. First, recall that one can use induction to show that 
\begin{equation}
[\hat p_r,\hat r^n]=-i\hbar n\hat r^{n-1}.
\label{eq: com_power}
\end{equation}
Using this, we immediately compute (for $m\ge 0$)
\begin{eqnarray}
[\hat p_r,J_m(k\hat r)]&=&\sum_{n=0}^\infty (-1)^n\frac{1}{n!(n+m)!}\left [\hat p_r,\left (\frac{k\hat r}{2}\right )^{2n+m}\right ]\nonumber\\
&=&-i\hbar\frac{k}{2}\sum_{n=0}^\infty (-1)^n\frac{2n+m}{n!(n+m)!}\left (\frac{k\hat r}{2}\right )^{2n+m-1}
\nonumber\\
&=&-i\hbar\frac{k}{2}\sum_{n=0}^\infty (-1)^n\frac{1}{n!(n+m-1)!}\left(\frac{k\hat r}{2}\right )^{2n+m-1}\nonumber\\
&-&i\hbar\frac{k}{2}\sum_{n=1}^\infty (-1)^n\frac{1}{(n-1)!(n+m)!}\left (\frac{k\hat r}{2}\right )^{2n+m-1}
\nonumber\\
&=&-i\hbar\frac{k}{2}\sum_{n=0}^\infty (-1)^n\frac{1}{n!(n+m-1)!}\left(\frac{k\hat r}{2}\right )^{2n+m-1}\nonumber\\
&+&i\hbar\frac{k}{2}\sum_{n=0}^\infty (-1)^n\frac{1}{n!(n+m+1)!}\left (\frac{k\hat r}{2}\right )^{2n+m+1}
\nonumber\\
&=&i\hbar \frac{k}{2} \left [-J_{m-1}(k\hat r)+J_{m+1}(k\hat r)\right ],
\label{eq: bessel_identity1}
\end{eqnarray}
where the second equality follows by evaluating the commutator according to Eq.~(\ref{eq: com_power}), the third by separating out $2n+m=(n+m)+n$, the fourth by shifting $n\rightarrow n+1$ in the second sum, and the final one via the definition of the Bessel functions in Eq.~(\ref{eq: bessel_def}).

The second identity we derive as follows:
\begin{eqnarray}
J_{m+1}(k\hat r)+J_{m-1}(k\hat r)&=&\sum_{n=0}^\infty(-1)^n\left (\frac{k\hat r}{2}\right )^{2n+m+1}\frac{1}{n!(n+m+1)!}\nonumber\\
&+&\sum_{n=0}^\infty (-1)^n\left (\frac{k\hat r}{2}\right )^{2n+m-1}\frac{1}{n!(n+m-1)!}
\nonumber\\
&=&-\sum_{n=1}^\infty(-1)^n\left (\frac{k\hat r}{2}\right )^{2n+m-1}\frac{1}{(n-1)!(n+m)!}\nonumber\\
&+&\sum_{n=0}^\infty (-1)^n\left (\frac{k\hat r}{2}\right )^{2n+m-1}\frac{1}{n!(n+m-1)!}
\nonumber\\
&=&\sum_{n=0}^\infty(-1)^n\left (\frac{k\hat r}{2}\right )^{2n+m-1}\frac{-n+n+m}{n!(n+m)!}\nonumber\\
&=&\frac{2m}{k\hat r}J_m(k\hat r),
\label{eq: bessel_identity2}
\end{eqnarray}
where in the second equality, we shifted the index by one in the first sum, combined all terms over the same denominator in the third equality, and used the definition of the Bessel function in the fourth.

Note that these Bessel functions are the ones that we use for the case where the energy is larger than the potential, because these Bessel functions oscillate and so does the wavefunction. There is another Bessel function, called the modified Bessel function of the second kind, and denoted $K_m$, which we use when the wavefunction exponentially decays. It is somewhat more challenging to try to derive the relationships between these Bessel functions without using any calculus, so we will simply state their fundamental identities. Their form is quite similar to the identities we derived for the Bessel functions of the first kind. These functions satisfy the following:
\begin{equation}
K_{-|m|}(\kappa \hat r) = K_{|m|}(\kappa \hat r)
\label{eq: bessel_k1}
\end{equation}
\begin{equation}
-K_{m-1}(\kappa \hat r)+K_{m+1}(\kappa \hat r) = \frac{2m}{\kappa \hat r}K_m(\kappa \hat r)
\label{eq: bessel_k2}
\end{equation}
\begin{equation}
[\hat p_r, K_m(\kappa \hat r)] = i\frac{\hbar\kappa}{2}[K_{m-1}(\kappa \hat r)+K_{m+1}(\kappa \hat r)].
\label{eq: bessel_k3}
\end{equation}
These relations are identical to those of the $J_m$ Bessel functions except for the changing of some signs.

\section{Radial momentum operator in two-dimensions}
We know that we can write any position eigenstate in two-dimensional coordinate space via Cartesian translation operators as
\begin{equation}
|x,y\rangle=e^{-\frac{i}{\hbar}(x\hat p_x+y\hat p_y)}|x{=}0,y{=}0\rangle,
\label{eq: 2d_state_cartesian}
\end{equation}
where $x$ and $y$ are numbers. When converting to polar coordinates, we let $x=r\cos\theta$ and $y=r\sin\theta$, so that $|r,\theta\rangle=\exp[-i(r\cos\theta\hat p_x+r\sin\theta\hat p_y)/\hbar]|r{=0},\theta{=}0\rangle$, where we use the convention that the origin ($r=0$) is at the angle $\theta=0$. Using the facts that
\begin{equation}
\hat p_r+i\frac{\hbar}{2\hat r}=\frac{\hat x}{\hat r}\hat p_x+\frac{\hat y}{\hat r}\hat p_y=\cos\hat\theta\hat p_x+\sin\hat\theta\hat p_y,
\label{eq: radial1}
\end{equation}
and
\begin{equation}
\hat p_\theta=-\frac{\hat y}{\hat r}\hat p_x+\frac{\hat x}{\hat r}\hat p_y=-\sin\hat\theta\hat p_x+\cos\hat\theta\hat p_y.
\label{eq: theta1}
\end{equation}
These two results immediately yield
\begin{equation}
e^{-\frac{i}{\hbar}(x\hat p_x+y\hat p_y)}=e^{-\frac{i}{\hbar}\left [ r\cos(\hat\theta-\theta)\left (\hat p_r+i\frac{\hbar}{2\hat r}\right )-r\sin(\hat\theta-\theta)\hat p_\theta\right ]}.
\label{eq: translate1}
\end{equation}
Using the fact that $\hat L_z$ commutes with both $\hat r$ and $\hat p_r$ allows us to use the identity
\begin{equation}
e^{-\frac{i}{\hbar}\theta\hat L_z}f(\hat\theta)e^{\frac{i}{\hbar}\theta\hat L_z}=
f\left ( e^{-\frac{i}{\hbar}\theta\hat L_z}\hat\theta e^{\frac{i}{\hbar}\theta\hat L_z}\right )=f(\hat\theta-\theta)
\label{eq: theta_identity}
\end{equation}
in Eq.~(\ref{eq: translate1}) to find
\begin{equation}
|x,y\rangle=e^{-\frac{i}{\hbar}\theta\hat L_z}e^{-\frac{i}{\hbar}\left [ r\cos\hat\theta\left (\hat p_r+i\frac{\hbar}{2\hat r}\right )-r\sin\hat\theta\hat p_\theta\right ]}|r{=}0,\theta{=}0\rangle.
\label{eq: translate2}
\end{equation}
Here, we used the fact that $\exp(i\theta\hat L_z/\hbar)|r{=}0,\theta{=}0\rangle=|r{=}0,\theta{=}0\rangle$ because rotating the state at the origin does not change it, so we can keep $\theta{=}0$.

We evaluate Eq.~(\ref{eq: translate2}) in a power series. First note that
\begin{equation}
[\hat\theta^n,\hat L_z]=i\hbar n\hat\theta^{n-1}
\end{equation}
allows us to derive that $[\sin\hat\theta,\hat L_z]=i\hbar\cos\hat\theta$ and hence $[\sin\hat\theta,\hat p_\theta]=i\hbar \cos(\hat\theta)/\hat r$. Using this result, we immediately compute
\begin{equation}
\left [r\cos\theta \left (\hat p_r+i\frac{\hbar}{2\hat r}\right )-r\sin\hat\theta\hat p_\theta\right ]|\theta{=}0\rangle=
r\left (\hat p_r-i\frac{\hbar}{2\hat r}\right )|\theta{=}0\rangle.
\end{equation}
Iterating this $n$ times for each term in the power series of the exponential gives us our final result
\begin{equation}
|x,y\rangle=|r,\theta\rangle=e^{-\frac{i}{\hbar}\theta\hat L_z}e^{-\frac{i}{\hbar}r\left (\hat p_r-i\frac{\hbar}{2\hat r}\right )}|r{=}0,\theta{=}0\rangle.
\label{eq: r0_state}
\end{equation}

\section{Operator identities for the three-dimensional case}

In this appendix, we show the details for how to work out powers of $(\hat p_r+i\hbar/\hat r)$ acting on $|\phi_{l{=}0}\rangle$. We start with the case of $r\le R$. Our methodology is to proceed by an inductive proof. We first work out the $n=1$ power:
\begin{equation}
\left (\hat p_r+i\frac{\hbar}{\hat r}\right )|\phi_{l{=}0}\rangle=i\hbar k\left (\frac{1}{k\hat r}-{\rm cotan}(k\hat r)\right )|\phi_{l{=}0}\rangle
\label{eq: 3d_n1}
\end{equation}
and the $n=2$ power:
\begin{eqnarray}
\left (\hat p_r+i\frac{\hbar}{\hat r}\right )^2|\phi_{l{=}0}\rangle&=&i\hbar k \left (\hat p_r+i\frac{\hbar}{\hat r}\right )\left (\frac{1}{k\hat r}-{\rm cotan}(k\hat r)\right )|\phi_{l{=}0}\rangle\nonumber\\
&=&(i\hbar k)^2\left (-1-\frac{2}{k\hat r}{\rm cotan}(k\hat r)+\frac{2}{k^2\hat r^2}\right )|\phi_{l{=}0}\rangle.
\label{eq: 3d_n2}
\end{eqnarray}
After determining a few more results explicitly, one can show that the general case is given by the following two results:
\begin{eqnarray}
&~&\left (\hat p_r+i\frac{\hbar}{\hat r}\right )^{2m+1}|\phi_{l{=}0}\rangle=
(i\hbar k)^{2m+1}\\
&\times& \sum_{n=0}^m(-1)^n\left [\frac{(2m+1)!}{(2n)!}\frac{1}{(k\hat r)^{2m-2n+1}}-\frac{(2m+1)!}{(2n+1)!}\frac{{\rm cotan}(k\hat r)}{(k\hat r)^{2m-2n}}\right ] |\phi_{l{=}0}\rangle\nonumber
\label{eq: 3d_induction1}
\end{eqnarray}
and
\begin{eqnarray}
&~&\left (\hat p_r+i\frac{\hbar}{\hat r}\right )^{2m+2}|\phi_{l{=}0}\rangle=
(i\hbar k)^{2m+2}\\
&\times&\left [ \sum_{n=0}^{m+1}(-1)^n\frac{(2m+2)!}{(2n)!}\frac{1}{(k\hat r)^{2m-2n+2}}- \sum_{n=0}^{m}(-1)^n\frac{(2m+2)!}{(2n+1)!}\frac{{\rm cotan}(k\hat r)}{(k\hat r)^{2m-2n+1}}\right ] |\phi_{l{=}0}\rangle\nonumber.
\label{eq: 3d_induction2}
\end{eqnarray}
We first verify that when $m=0$, these two results reproduce those in Eqs.~(\ref{eq: 3d_n1}) and (\ref{eq: 3d_n2}).
Then we assume it holds for all $n$ up to $2m$ and prove it holds for $2m+1$. We compute
\begin{eqnarray}
&~&\left (\hat p_r+i\frac{\hbar}{\hat r}\right )^{2m+1}|\phi_{l{=}0}\rangle=
(i\hbar k)^{2m}\left (\hat p_r+i\frac{\hbar}{\hat r}\right )\\
&\times&\left [ \sum_{n=0}^{m}(-1)^n\frac{(2m)!}{(2n)!}\frac{1}{(k\hat r)^{2m-2n}}- \sum_{n=0}^{m-1}(-1)^n\frac{(2m)!}{(2n+1)!}\frac{{\rm cotan}(k\hat r)}{(k\hat r)^{2m-2n-1}}\right ] |\phi_{l{=}0}\rangle\nonumber\\
&=&(i\hbar k)^{2m+1}\left [  \sum_{n=0}^{m}(-1)^n\left \{ \frac{(2m)!(2m-2n)}{(2n)!}\frac{1}{(k\hat r)^{2m-2n+1}}\right .- \frac{(2m)!}{(2n)!}\frac{{\rm cotan}(k\hat r)}{(k\hat r)^{2m-2n}}\right .\nonumber\\
&~&~~~~~~+\left . \frac{(2m)!}{(2n)!}\frac{1}{(k\hat r)^{2m-2n+1}}\right \}\nonumber\\
&-& \sum_{n=0}^{m-1}(-1)^n\left \{\frac{(2m)!}{(2n+1)!}\frac{{\rm cosec}^2(k\hat r)}{(k\hat r)^{2m-2n-1}}+\frac{(2m)!(2m-2n-1)}{(2n+1)!}\frac{{\rm cotan}(k\hat r)}{(k\hat r)^{2m-2n}}\right \}\nonumber\\
&+&\left . \sum_{n=0}^{m-1}(-1)^n\left \{ \frac{(2m)!}{(2n+1)!}\frac{{\rm cotan}^2(k\hat r)}{(k\hat r)^{2m-2n-1}}-\frac{(2m)!}{(2n+1)!}\frac{{\rm cotan}(k\hat r)}{(k\hat r)^{2m-2n+1}}\right \}\right ] |\phi_{l{=}0}\rangle\nonumber\\
&=&(i\hbar k)^{2m+1}\sum_{n=0}^m(-1)^n\left [\frac{(2m+1)!}{(2n)!}\frac{1}{(k\hat r)^{2m-2n+1}}-\frac{(2m+1)!}{(2n+1)!}\frac{{\rm cotan}(k\hat r)}{(k\hat r)^{2m-2n}}\right ]|\phi_{l{=}0}\rangle,\nonumber
\end{eqnarray}
where we used a number of identities including the following: ${\rm cosec}^2(x)-{\rm cotan}^2(x)=1$, we extended the sum that ranged up to $m-1$ to $m$ for the ${\rm cotan}(x)$ term, because the term with $n=m$ vanished, while we shifted the sum for the inverse power of $\hat r$ by $n\rightarrow n+1$ and then added on the term with $n=0$ because it vanished. The final result is the one we wanted to show in Eq.~(\ref{eq: 3d_induction1}).

Now, we assume the identity holds up to $m+1$ and prove the result for $m+2$:
\begin{eqnarray}
&~&\left (\hat p_r+i\frac{\hbar}{\hat r}\right )^{2m+2}|\phi_{l{=}0}\rangle=
(i\hbar k)^{2m+1}\left (\hat p_r+i\frac{\hbar}{\hat r}\right )\\
&\times&\sum_{n=0}^m(-1)^n\left [\frac{(2m+1)!}{(2n)!}\frac{1}{(k\hat r)^{2m-2n+1}}- \frac{(2m+1)!}{(2n+1)!}\frac{{\rm cotan}(k\hat r)}{(k\hat r)^{2m-2n}}\right ] |\phi_{l{=}0}\rangle\nonumber\\
&=&(i\hbar k)^{2m+2}\sum_{n=0}^m(-1)^n\left [  \left \{ \frac{(2m+1)!(2m-2n+1)}{(2n)!}\frac{1}{(k\hat r)^{2m-2n+2}}\right .\right .\nonumber\\
&~&~~~- \frac{(2m+1)!}{(2n)!}\frac{{\rm cotan}(k\hat r)}{(k\hat r)^{2m-2n+1}}+ \frac{(2m+1)!}{(2n)!}\frac{1}{(k\hat r)^{2m-2n+2}}\nonumber\\
&~&~~~- \frac{(2m+1)!}{(2n+1)!}\frac{{\rm cosec}^2(k\hat r)}{(k\hat r)^{2m-2n}}-\frac{(2m+1)!(2m-2n)}{(2n+1)!}\frac{{\rm cotan}(k\hat r)}{(k\hat r)^{2m-2n+1}}\nonumber\\
&~&~~~+\left . \left . \frac{(2m+1)!}{(2n+1)!}\frac{{\rm cotan}^2(k\hat r)}{(k\hat r)^{2m-2n}}-\frac{(2m+1)!}{(2n+1)!}\frac{{\rm cotan}(k\hat r)}{(k\hat r)^{2m-2n+1}}\right \}\right ] |\phi_{l{=}0}\rangle\nonumber\\
&=&(i\hbar k)^{2m+1}\left [\sum_{n=0}^{m+1}(-1)^n\frac{(2m+2)!}{(2n)!}\frac{1}{(k\hat r)^{2m-2n+2}}\right .\nonumber\\
&~&~~~~~~~~~~~~\left .-\sum_{n=0}^m(-1)^n\frac{(2m+2)!}{(2n+1)!}\frac{{\rm cotan}(k\hat r)}{(k\hat r)^{2m-2n+1}}\right ]|\phi_{l{=}0}\rangle,\nonumber
\end{eqnarray}
which again is what we needed to establish Eq.~(\ref{eq: 3d_induction2}).

We now need to evaluate the overlap of these expressions with the bra $\langle r{=}0|$. At first, it looks like we will obtain a divergence, but we will find all negative powers of $\hat r$ will vanish and we will be left with finite results.
We start with the odd powers. We find it is easier to evaluate the matrix element inserting the factor $1=\sin(k\hat r)/\sin(k\hat r)$, which yields
\begin{eqnarray}
&~&\langle r{=}0|\frac{\sin(k\hat r)}{\sin(k\hat r)}\left (\hat p_r+i\frac{\hbar}{\hat r}\right )^{2m+1}|\phi_{l{=}0}\rangle=\langle r{=}0|\frac{1}{\sin(k\hat r)}(i\hbar k)^{2m+1}\nonumber\\
&\times& \sum_{n=0}^m(-1)^n\left [\frac{(2m+1)!}{(2n)!}\frac{\sin(k\hat r)}{(k\hat r)^{2m-2n+1}}-\frac{(2m+1)!}{(2n+1)!}\frac{\cos(k\hat r)}{(k\hat r)^{2m-2n}}\right ] |\phi_{l{=}0}\rangle\nonumber\\
&=&\langle r{=}0|\frac{(i\hbar k)^{2m+1}}{\sin(k\hat r)}\sum_{n=0}^m\sum_{n'=0}^\infty(-1)^{n+n'}\frac{(2m+1)!}{(2n)!(2n')!}\left (\frac{1}{2n'+1}-\frac{1}{2n+1}\right )\nonumber\\
&~&~~~~\times(k\hat r)^{2n'-2m+2n}|\phi_{l{=}0}\rangle=0.
\end{eqnarray}
We derived this result by simply expanding the $\sin$ and $\cos$ in their power series. The final result vanishes because in every case where the exponent of the $(k\hat r)$ term is negative or zero, the term with a specific $n$ value is canceled by a term with the same $n'$ value. Hence, the lowest-order contribution goes like $(k\hat r)^2$. But the limit as $r\to 0$ of $r^2/\sin(kr)$ is zero. This proves that all odd powers vanish.

For the even powers, we have 
\begin{eqnarray}
&~&\langle r{=}0|\frac{\sin(k\hat r)}{\sin(k\hat r)}\left (\hat p_r+i\frac{\hbar}{\hat r}\right )^{2m+2}|\phi_{l{=}0}\rangle=\langle r{=}0|\frac{1}{\sin(k\hat r)}(i\hbar k)^{2m+2}\nonumber\\
&\times& \left [\sum_{n=0}^{m+1}(-1)^n\frac{(2m+2)!}{(2n)!}\frac{\sin(k\hat r)}{(k\hat r)^{2m-2n+2}}-\sum_{n=0}^m(-1)^n\frac{(2m+2)!}{(2n+1)!}\frac{\cos(k\hat r)}{(k\hat r)^{2m-2n+1}}\right ]|\phi_{l{=}0}\rangle,\nonumber\\
&=&\langle r=0|\frac{(i\hbar k)^{2m+2}}{\sin(k\hat r)}\left [ \sum_{n=0}^m\sum_{n'=0}^\infty(-1)^{n+n'}\frac{(2m+2)!}{(2n)!(2n')!}\left (\frac{1}{2n'+1}-\frac{1}{2n+1}\right )\right .\nonumber\\
&~&\times(k\hat r)^{2n'-2m+2n-1}-(-1)^{m+1}\sin(k\hat r)\Biggr ]|\phi_{l{=}0}\rangle=\frac{(\hbar k)^{2m+2}}{2m+3}\langle r=0|\phi_{l{=}0}\rangle.
\end{eqnarray}
Once again, all negative powers of $(k\hat r)$ vanish due to the cancellations in the double summations. The terms that are linear in $(k\hat r)$ from the double sum have $n=0$ and $n'=m+1$ and then the last term cancels the $\sin$ terms in the numerator and denominator. This then yields the final result.

The last derivation we need to do is to evaluate the matrix element of the powers when $r>R$. Here, the wavefunction satisfies $\hat p_r|\phi_{l{=}0}\rangle=i\hbar\kappa|\phi_{l{=}0}\rangle$. We define $f_n(\kappa \hat r)$ via
\begin{equation}
\left ( \hat p_r+i\frac{\hbar}{\hat r}\right )^n|\phi_{l{=}0}\rangle=(i\hbar\kappa)^nf_n(\kappa\hat r)|\phi_{l{=}0}\rangle,
\end{equation}
so that $f_1(\kappa\hat r)=1+1/(\kappa\hat r)$. One can immediately verify that
\begin{equation}
f_n(\kappa\hat r)=\frac{1}{i\hbar\kappa}\left [\hat p_r,f_{n-1}(\kappa\hat r)\right ]+f_{n-1}(\kappa\hat r)\left (1+\frac{1}{\kappa\hat r}\right ).
\label{eq: 3d_induction5}
\end{equation}
This recurrence relation can be solved by substituting in a degree-$n$ polynomial for $f_n$. We find that
\begin{equation}
f_n(\kappa\hat r)=\sum_{m=0}^n\frac{n!}{(n-m)!}\frac{1}{(\kappa\hat r)^m}.
\end{equation}
One can directly verify that this form satisfies Eq.~(\ref{eq: 3d_induction5}). This relation is then used in the evaluation of the wavefunction for $r>R$ above.

\end{document}